# Medical Imaging and Computational Image Analysis in COVID-19 Diagnosis: A Review

Shahabedin Nabavi[1], Azar Ejmalian[2], Mohsen Ebrahimi Moghaddam[1], Ahmad Ali Abin[1], Alejandro F. Frangi[3], Mohammad Mohammadi[4,5], Hamidreza Saligheh Rad[6]

1- Faculty of Computer Science and Engineering, Shahid Beheshti University, Tehran, Iran.
2- Anesthesiology Research Center, Shahid Beheshti University of Medical Sciences, Tehran, Iran.
3- Centre for Computational Imaging and Simulation Technologies in Biomedicine (CISTIB), School of Computing, University of Leeds, Leeds, UK.
4- Department of Medical Physics, Royal Adelaide Hospital, Adelaide, South Australia, Australia.
5- School of Physical Sciences, The University of Adelaide, Adelaide, South Australia, Australia.
6- Quantitative MR Imaging and Spectroscopy Group (QMISG), Tehran University of Medical Sciences, Tehran, Iran.

**Abstract:**

Coronavirus disease (COVID-19) is an infectious disease caused by a newly discovered coronavirus. The disease presents with symptoms such as shortness of breath, fever, dry cough, and chronic fatigue, amongst others. Sometimes the symptoms of the disease increase so much they lead to the death of the patients. The disease may be asymptomatic in some patients in the early stages, which can lead to increased transmission of the disease to others. Many studies have tried to use medical imaging for early diagnosis of COVID-19. This study attempts to review papers on automatic methods for medical image analysis and diagnosis of COVID-19. For this purpose, PubMed, Google Scholar, arXiv and medRxiv were searched to find related studies by the end of April 2020, and the essential points of the collected studies were summarised. The contribution of this study is four-fold: 1) to use as a tutorial of the field for both clinicians and technologists, 2) to comprehensively review the characteristics of COVID-19 as presented in medical images, 3) to examine automated artificial intelligence-based approaches for COVID-19 diagnosis based on the accuracy and the method used, 4) to express the research limitations in this field and the methods used to overcome them. COVID-19 reveals signs in medical images can be used for early diagnosis of the disease even in asymptomatic patients. Using automated machine learning-based methods can diagnose the disease with high accuracy from medical images and reduce time, cost and error of diagnostic procedure. It is recommended to collect bulk imaging data from patients in the shortest possible time to improve the performance of COVID-19 automated diagnostic methods.

**List of abbreviations:**

AI: Artificial Intelligence

COVID-19: Coronavirus disease 2019

CT: Computed Tomography

CXR: Chest X-ray

PCR: Polymerase Chain Reaction

RT-PCR: real-time Reverse transcription-PCR

NAAT: Nucleic acid amplification test

GGO: Ground-glass opacity

FDG: FluoroDeoxyGlucose

PET: Positron Emission Tomography

SPECT: Single-photon Emission Computed Tomography

MRI: Magnetic Resonance Imaging

CNN: Convolutional Neural Network

BCNN: Bayesian CNN

AE: Auto-encoder

GAN: Generative Adversarial Network

CGAN: Conditional GAN

SVM: Support Vector Machine

AUC: Area Under the Curve

GLCM: Grey-Level Co-occurrence Matrix

LDP: Local Directional Pattern

GLRLM: Grey-Level Run Length Matrix

GLSZM: Grey-Level Size Zone Matrix

DWT: Discrete Wavelet Transform

RF: Random Forest

SNR: Signal-to-Noise Ratio

CNR: Contrast-to-Noise Ration

PPV: Positive Predictive Value

DECAPS: Detail-Oriented Capsule Networks

ICU: Intensive Care Unit

SMOTE: Synthetic Minority Over-Sampling Technique

## 1- Introduction

Coronaviruses are a large family of viruses that cause disease in humans in the form of a common cold to more severe respiratory infections. An infectious disease caused by a newly discovered coronavirus, also known as COVID-19, is a disease that causes an acute respiratory syndrome, which can lead to the death of infected patients. The disease was first seen in December 2019 in Wuhan, China, and eventually became a global pandemic. According to the official statistics, the number of people infected with the disease had reached over 16 million worldwide, with over 600,000 deaths until end of July 2020 [1].

Patients have a variety of symptoms during the illness, including shortness of breath, fever, dry cough, and chronic fatigue. Sometimes the symptoms are so severe in patients they can be fatal. The leading cause of transmission is the contact of the person's hand with the contaminated surfaces and then touching the face. Despite many efforts by scientists, there is currently no vaccine for the disease. Therefore, the main advice to prevent infection is to observe personal hygiene by regularly washing hands, disinfecting surfaces and covering the airways with a mask [2].

CT imaging has been proposed as one way to diagnose the disease. A large number of studies have been published on the role of medical imaging in the diagnosis of this disease in the short time since the outbreak stage. Many researchers in medical image analysis are also seeking to provide artificial intelligence (AI) based solution for the automatic diagnosis of the disease based on medical images. This review provides a summary of peer-reviewed research articles, conference papers, case reports and letters to the journal editors related to the role of imaging and also peer-reviewed research articles, conference papers and preprints related to medical image analysis in COVID-19 to help other researchers in conducting their studies due to the importance of this disease and its destructive effects on societies. PubMed, Google Scholar, arXiv and medRxiv were searched for all peer-reviewed journal articles, conference papers, case reports, letters to the journal editors and preprints that fulfil the following selection criteria by the end of April 2020:

- the role of all different medical imaging modalities in COVID-19 diagnosis;
- imaging findings related to COVID-19;
- advice and statements for using imaging in COVID-19 diagnosis;
- automated methods for detection and classification COVID-19 based on medical imaging data.

The rest of this review is structured as followed. Section 2 describes the contributions of the role of medical imaging in COVID-19 diagnosis. In this section, many articles, case reports and letters to editors related to this field are reviewed so the reader can understand the main points of these articles can pick up with literature and critical contributions in a short time. In section 3, articles related to automatic methods in the detection of COVID-19 based on AI techniques are reviewed. Finally, in section 4, this article concludes with a discussion and an outlook for future studies.

## 2- The role of imaging in COVID-19 diagnosis

There are several ways to detect COVID-19, including real-time reverse transcription-polymerase chain reaction (RT-PCR) and the nucleic acid amplification test (NAAT). Because often the test results may be negative despite having the person infected, and asymptomatic infections can spread the infection, there is a need for a more careful approach to diagnosis. Some studies and reports confirm that medical imaging can be an effective way to diagnose COVID-19 infection, even if the patient is asymptomatic. Therefore, where it is impossible to access the above tests, medical imaging can help diagnose the disease and prevent its spread in asymptomatic patients. In this section, we review studies that have been published on the role of medical imaging in the diagnosis of COVID-19. We summarise the most essential points of view of articles, including the features of the disease in medical images.

In the study of Rodrigues et al. [3], chest imaging findings were examined in the literature. Some pointed out that decisions should be made regarding determining medical imaging as a screening tool in patients with different severity of the disease. In the study of Nair et al. [4], the use of CT imaging in the diagnosis and management of COVID-19 disease in the UK was investigated by asking several questions and answering them. This study considers the results of CT imaging as a standard in the diagnosis of COVID-19 to be contradictory. As another example, in the study of Huang et al. [5], it is emphasised that CT imaging should not be recommended as a screening tool in the early diagnosis of COVID-19 due to two issues. The first is failure to prove that CT imaging can always succeed in diagnosing COVID-19, and the second is that overexposure of patients to radiation can have long-term adverse effects. Raptis et al. in their review study [6] concluded that the studies that have examined the role of CT imaging in the diagnosis and management of COVID-19 could not prove this role due to their limitations. Although some studies have debunked the role of medical imaging in the diagnosis of COVID-19, many studies have attempted to prove its importance by expressing the imaging features of the disease, and using CT imaging is considered an effective solution in early diagnosis, severity assessment and patient management in the progression of COVID-19 [7-18].

In the study of Zhu et al. [19], 34 studies with 4121 COVID-19 patients have been systematically reviewed. The main features of CT in this study are ground-glass opacities (GGOs), air bronchogram sign, crazy-paving pattern, consolidation, pleural thickening, lymphadenopathy, and pleural effusion, respectively. Also, Lesion shapes are patchy, spider web sign, cord-like, and nodular. In the study of Li [20], the importance of using chest CT for the diagnosis and follow-up of COVID-19 patients has been investigated. CT features at different stages of the disease are different according to previous studies. Hani et al. [21], CT imaging has been cited as a critical complement in the diagnosis of COVID-19. CT features have been described as peripheral GGOs with multifocal distribution, and a progressive evolution towards organizing pneumonia patterns.

In another systematic review study conducted by Bao et al. [22], 13 studies have been reviewed. CT characteristics have been listed as GGO, GGO with mixed consolidation, adjacent pleura thickening, interlobular septal thickening, air bronchograms, crazy-paving pattern, pleural effusion, bronchiectasis, pericardial effusion, and lymphadenopathy, respectively. The most anatomic distributions are bilateral lung infection and peripheral distribution. In the study by Salehi et al. [23], 30 studies consisting of 19 case series and 11 case reports with a total of 919 patients were systematically reviewed. CT findings are included GGO, bilateral involvement, peripheral distribution, and multilobar involvement. Other CT findings include interlobular septal thickening, bronchiectasis, pleural thickening, and subpleural involvement. Rarely found findings include pleural effusion, pericardial effusion, lymphadenopathy, cavitation, CT halo sign, and pneumothorax.

**2-1- Computed Tomography**

CT imaging has been widely used as a fundamental modality in the diagnosis of COVID-19 in the studies. High-resolution [24-29], low-dose [30, 31], thin-section [32, 33] and spiral [34] CT imaging are mentioned as a main modality in some of researches.

Some studies have shown that CT imaging is insufficient or incapable as a diagnostic modality for COVID-19 [35-38]. Some other studies show misdiagnosis in early-stage patients of COVID-19 [25], and propose the combination of CT imaging and clinical findings for better diagnosis of COVID-19 [39, 40], especially in children [41-43]. A large number of studies have reported the importance of CT imaging in the diagnosis of COVID-19, and CT features related to the patients infected with COVID-19.

GGOs, patchy and wedge-shaped GGOs, consolidation, vascular enlargement and thickening, interlobular septal thickening, interstitial thickening, air bronchogram sign, fibrotic lesions, pleural effusion, crazy-paving pattern, linear and rounded opacities, reticulation, fine reticular opacity, subpleural and central lesions, irregular solid nodules, interstitial pulmonary oedema, halo sign, reversed-halo sign, architectural distortion, bronchial wall thickening, subpleural bands, traction bronchiectasis, intrathoracic lymph node enlargement, lymphadenopathy, thickening of the adjacent pleura, cystic changes, cord-like lesions, thickening of the bronchovascular bundles, pleural thickening, cavitation, tree-in-bud sign, interlobar fissure displacement, pericardial effusion, concomitant hydropericardium and/or hydrothorax, thickened lobular septum, thickened bronchial wall, vacuolar sign, bronchiolar dilatation, secondary tuberculosis, paving stone sign, pleural retraction sign, fine mesh shadow, pneumatocele, spider web sign, enlarged mediastinal nodes, underlying pulmonary emphysema, bullae of lung and obsolete tuberculosis and thickened leaflet interval are CT findings of patients with COVID-19 in the collected studies (Refer to Table 1).

Some studies have individually examined the characteristics of CT in children. A study by Li et al. [44], which was performed on five children, has found patchy GGOs as the main characteristic of CT in children with COVID-19 and believed that the abnormalities in CT images of children are similar but milder than those of adults. The study of Zhu et al. [45], which was performed on 44 younger (47.5 ± 8.7 y old) and 28 older patients (68.4 ± 6.0 y old) with COVID-19, despite the reporting of some similar CT features among younger and older patients, considers it more likely that extensive lung lobe involvement, subpleural line and pleural thickening will occur in older patients. The study of Feng et al. with 15 cases of paediatric patients diagnosed with COVID-19 [46], identified small nodular and speckled GGOs as the main features in CT images of these patients. Another study also described CT findings in children as milder than in adults [27]. In a study of nine children with COVID-19 aged 0 to 3 years by Zhou et al. [43], the CT findings are nodular lesions, patchy lesions, GGO with consolidation and halo sign noted to be milder than in adults. In their study, Liu et al. [41] believe that a history of exposure and clinical symptoms may be more useful in the screening of COVID-19 in children than CT imaging. Procalcitonin elevation and consolidation with surrounding halo signs are frequent in paediatric-patients than adults based on the study of Xia et al. [42].

Studies have also been performed in pregnant women with COVID-19. According to Wu et al. [31], the CT findings in pregnant women are similar to those in non-pregnant women. In the study by Liu et al. [41], consolidation has been described as a more common CT feature in pregnant women.

Table 1 is an overview of studies that have examined the role of CT imaging in the diagnosis of COVID-19. These studies have five or more cases of COVID-19, which are sorted based on the number of cases. CT and other related findings are summarised for each study. Table 2 deals with studies on CT imaging for COVID-19 diagnosis with less than five cases mostly written as a case report or letter to the editor. In these tables, to acquaint researchers with the studies conducted in this field, the number of cases, the features of COVID-19 infection in the studied images and the findings or essential points of these studies are summarised.

**2-2- Chest X-Ray**

Among the reviewed studies, seven studies have focused on the signs of COVID-19 in CXR images. In the study of Jacobi et al. [47], which is a pictorial review, the possibility of using portable CXR imaging in the diagnosis of COVID-19 has been investigated due to its availability in most medical centres. Based on this research, CXR imaging also provides the ability to detect COVID-19. Zhang et al. have reported that the combination of clinical features and radiological findings can predict the severity of COVID-19 [40]. In their study [48], Wang et al. believe that since the diagnostic role of CT imaging has not been accurately proven, it is better to use a modality such as CXR with less radiation, especially for children.

In the study of Lomoro et al. [49], which has been performed on 58 patients with COVID-19, CXR manifestations are consolidation and hazy increased opacity. In the study of Wong et al. [50], conducted with 64 patients with COVID-19, GGOs, consolidation and pleural effusion have been reported as CXR findings. The results of this study show bilateral lower zone consolidation, which peaked at 10-12 days from symptom onset. CXR findings also had sensitivity lower than initial RT-PCR testing. CXR manifestations are parenchymal abnormalities, consolidation, GGOs, single nodular opacity and patchy opacities in the study of Yoon et al. [51]. According to this study, a large proportion of patients with COVID-19 have normal CXR images. Table 3 summarizes the studies related to this section.

**2-3- Ultrasound**

Few studies have suggested the use of lung ultrasound to detect COVID-19. In Lu et al. (2020) [52], lung ultrasound signs are interstitial pulmonary oedema and pulmonary consolidations. The study concluded that although the lung ultrasound diagnostic efficacy on the detection of COVID-19 in mild and moderate patients is relatively low, it is high in severe patients. Thirty patients were examined in this study. In Lomoro et al. (2020) [49], lung ultrasound signs are B-lines patterns (focal, multifocal, and confluent) due to interlobular septal thickening or hazy opacities, subpleural consolidation, thickened pleural line, pleural effusion and mixed pattern with A- and B-lines based on information from 58 COVID-19 patients.

In the study of Wang et al. [48], lung ultrasound is considered a suitable modality for the diagnosis of COVID-19 due to its non-radiation nature. Vetrugno et al. [53] also believe that lung ultrasound can show the severity and involvement of the lung in COVID-19. Kalafat et al. (2020) [54] is a case report of positive lung ultrasound findings of COVID-19 in a pregnant woman with an initially negative RT-PCR result.

**2-4- $^{18}$F-FDG PET/CT**

Some studies have confirmed the role of this modality and its sensitivity to the detection of COVID-19. Based on Lutje et al. (2020) [55], this modality can play a complementary role in the management of COVID-19. It means this modality has the potential to diagnose COVID-19, and it can be used for estimating the extent to which organs are involved and determining the response to treatment in patients. Deng et al. [56] also highlighted the sensitivity of this modality in COVID-19 diagnosis and monitoring disease progression or the success rate of treatment. In a study by Qin et al. [57] that described the results of $^{18}$F-FDG PET/CT for four patients, all patients had peripheral GGOs and consolidations in over two pulmonary lobes. According to this study, although it is impossible to use this modality widely, it has good potential in detecting the complex cases of COVID-19. Zou and Zhu [58] and Polverari et al. [59] are case reports in which imaging was performed with this modality. The imaging results revealed bilateral, diffuse, and intense FDG uptake in the lower lobes and less intense uptake in the remaining lobes.

However, there are also conflicting points about the effectiveness of this modality. Prolonged imaging may cause the disease to spread in imaging centres [60]. Also, due to the large number of COVID-19 patients, the $^{18}$F-FDG PET/CT imaging capacity cannot cover this number of patients [61].

**2-5- Other modalities:**

Some other studies have talked about other imaging modalities to detect COVID-19. Tulchinsky et al. [62] suggests that since CT-SPECT can diagnose COVID-19, nuclear medicine physicians should be familiar with the features of the disease in the images. The study of Poyiadji et al. [63] is a case report related to using MRI images to detect COVID-19–associated acute necrotising haemorrhagic encephalopathy. Acute necrotising encephalopathy is a rare complication of viral infections like influenza.

When the level of D-dimer increases during hospitalisation or sudden clinical deterioration, CT angiography can be a life-saving option for patients, as patients with COVID-19 may be associated with acute pulmonary embolism [64].

## 3- Automated image analysis methods for COVID-19 diagnosis

Due to challenges such as the unavailability of PCR testing in all centres of COVID-19 and the high false-negative rate of this test [65], which has been mentioned in many studies in this field, medical imaging for early detection of COVID-19 has received more attention. However, evaluating many of medical images in the epidemic situations will undoubtedly be a time-consuming and error-prone process. Therefore, given the advances in machine learning, more reliance on these techniques for the automatic diagnosis of COVID-19 based on medical images should be considered [66-70].

AI-based methods can provide automated tools for detecting COVID-19 [69]. The distinguishing features must first be extracted from the image to create automated diagnosis methods. The feature extraction process can either be based on handcrafted feature extraction methods or deep learning approaches [71]. Machine learning approaches can then be used for medical image classification, medical image segmentation, severity assessment of disease and other possible tasks based on extracted features.

This section presents an overview of automated methods in the diagnosis of COVID-19 based on medical imaging. We review the methods, the main contributions of the studies and the imaging datasets available in this field. We also discuss the performance of the methods.

### 3-1- Deep-learning-based approaches

With deep learning, many operations required to analyse images and extract features from images have become more manageable. Convolutional neural networks (CNNs) [72] have been widely used for image analysis, and this review cites several studies that use this approach to detect COVID-19 from CT or CXR images. A look at the famous CNN architectures shows that they all consist of three types of layers. These layers include convolutional, pooling, and fully-connected layers. Convolutional layers, which are based on the use of convolution kernels, are responsible for extracting features from images. Pooling layers reduce the resolution of feature maps based on operations such as average or max-pooling so that they can achieve shift-invariance. Fully-connected layers aim to perform classification based on obtained feature maps from previous layers. The reason for using multiple layers is that the kernels of the first convolutional layer are used to extract the low-level image features such as edges, and the subsequent convolutional layers extract the high-level features of the image. Softmax operation is usually used for the final classification, while other methods such as Support Vector Machine (SVM) can also be used for this purpose [73].

There is a need for large-scale data to take advantage of deep learning approaches, and some studies have collected data sets required to evaluate the automatic methods of detecting COVID-19 [74, 75]. Big datasets are not yet available for deep learning methods because it has not been long since the COVID-19 pandemic. Therefore, many studies have addressed the challenge of data scarcity using data augmentation [76] or transfer learning [77].

Transfer learning is a way in which knowledge gained from one domain can learn in another domain. It means it is possible to train a deep neural network and store the knowledge obtained on a domain where there is enough data and that knowledge can be used to train the network with little data from another domain. Two different strategies of transfer learning can be used for image classification. In one strategy, the pre-trained network can be used as a feature extractor, and in another strategy, the pre-trained network can be fine-tuned on images of COVID-19 patients. There are contradictory results regarding the use of these strategies, but in general, using transfer learning dramatically improves the classification accuracy. Using this method can sometimes even outperform human experts [78].

In the study of Varshni et al. [79], the use of pre-trained CNNs for feature extraction of CXR images along with different classifiers to distinguish between normal and abnormal images has been investigated for pneumonia detection. Some pre-trained CNN models including Xception [80], VGG16 and VGG-19 [81], ResNet-50 [82], DenseNet-121 and DenseNet-169 [83], and classifiers including Random Forest (RF), K-nearest neighbours, Naive Bayes and SVM have been evaluated in the study. Based on statistical results, the combination of DenseNet-169 for the feature extraction and SVM for the classification has been selected, and the accuracy of this proposed method is 80%. In the study of Narin et al. [84], transfer learning has been used on three deep CNNs including ResNet50, InceptionV3 and InceptionResNetV2. A total of 100 CXR images, including 50 images of patients with COVID-19 and 50 normal images, were used to learn the networks, and the results show a 98% accuracy of ResNet50 using 5-fold cross-validation. Another similar study [85], which used the transfer learning technique to diagnose COVID-19 automatically, evaluates seven deep convolutional neural networks. A 90% accuracy for VGG19 and DenseNet201 in a dataset, including 50 normal images and 25 images of COVID-19 patients, has been achieved. In the study of Ghoshal and Tucker [86], a pre-trained network called ResNet50V2 has been used. Dropweights based Bayesian CNN (BCNN) has also been used to estimate uncertainty in deep learning strategies to improve diagnostic performance. The study, with 5,941 CXR images, including 68 images of patients with COVID-19, achieved an accuracy of about 89%.

Data augmentation is a way to address the problem of data limitation to avoid network overfitting. Data augmentation can be done with basic image processing techniques or deep learning approaches. The former includes geometric and lightening transformations, image mixing and filtering. Deep learning approaches include generative adversarial learning [76].

Khalifa et al. [87] have presented a combined approach involving transfer learning and the use of generative adversarial networks (GAN) for data augmentation. A 99% accuracy has been achieved using Resnet18 on a dataset including 5863 CXR images. Loey et al. [88] have used conditional GAN (CGAN) for data augmentation, and this method has improved the performance of classification. In their study, Farooq and Hafeez [89] have used methods such as vertical flips, random rotation of images and lighting condition transformations for data augmentation. Similarly, the study of Apostolopoulos et al. [90] has used random rotation and random horizontal and vertical shift towards any direction for data augmentation. Data augmentation has been done in the study of Zheng et al. [91] using random affine transformation and colour jittering. The affine transformation comprised rotation, horizontal and vertical translations, scaling and shearing in the width dimension. The colour jittering adjusted brightness and contrast. Hu et al. [92] have augmented the data by cropping square patches at the centre of the input frames, rotation with a random angle, random horizontal reflection and contrast adjustment using randomly darkening or brightening.

Some studies have also tried to present a deep neural network architecture from scratch. In a study by Wang et al. [93], a deep convolutional neural network called COVID-NET is proposed to detect COVID-19 based on 13,975 collected CXR images. This study compared the results of the proposed method with the VGG-19 and ResNet-50, which shows that the accuracy of 93.3% for the proposed method is superior to the other methods. As another example, Oh et al. [94] identified COVID-19 in their study based on a patch-based CNN approach with a relatively small number of trainable parameters.

Table 4 summarises the automated deep learning-based approaches for COVID-19 diagnosis.

**3-2- Other approaches:**

Some studies use non-deep learning methods to detect COVID-19 automatically. Barstugan et al.'s study [95] used methods such as grey-level co-occurrence matrix (GLCM), local directional pattern (LDP), grey-level run length matrix (GLRLM), grey-level size zone matrix (GLSZM), and discrete wavelet transform (DWT) to extract the feature from 150 CT images. SVM has also been used for classification, and the results show a 99.68% accuracy of the classification using 10-fold cross-validation and GLSZM feature extraction method. In the study of Tang et al. [96], 63 quantitative features along with a RF model have been used to assess the severity of COVID-19 based on CT images of 176 patients infected with COVID-19. The accuracy of 87.5% has been obtained based on three-fold cross-validation. In the study of Al-Karawi et al. [97], FFT-Gabor scheme and SVM have been used for feature extraction and classification respectively, with results showing a 95.37% accuracy rate among 275 positive COVID-19, and 195 negative patients.

## 4- Discussion

### 4-1- Overview

In this review study, many articles and preprints related to the role of medical imaging and automatic methods of medical image analysis in the diagnosis of COVID-19 were examined. Despite some articles that deny the role of medical imaging in the diagnosis and management of COVID-19, many studies have highlighted this role and examined the characteristics of the disease in medical images. Despite the sometimes contradictory results, a significant portion of the articles emphasises the use of medical images including CT, CXR, ultrasound, $^{18}$F-FDG PET/CT and so on to diagnose COVID-19. The reasons for these studies are that the disease shows visible signs in medical images that can be used for early detection of COVID-19 in the lack of access to RT-PCR and other related methods.

Efforts have also been made to diagnose COVID-19 automatically from CT and CXR images using machine learning techniques. The wide range of machine learning methods, especially deep learning, can be used for COVID-19 diagnosis.

### 4-2- Key aspects of medical imaging for COVID-19 diagnosis

CT is the primary modality in early detection of COVID-19 because a significant portion of the reviewed studies, including 122 studies, examined the role of CT imaging. GGO and consolidation are the most common COVID-19 features in CT images based on the significant number of studies. The study of Ai et al. [98] with 1014 patients studied has the highest population compared to other studies. After that, Zhang et al. [40] and Ling et al. [36] studies are in the next ranks with 645 and 295 patients, respectively. The number of patients in these studies shows there are limitations in terms of patient information. Therefore, more comprehensive studies are needed. There have also been studies on children and pregnant women. Given the conflicting results, more studies are needed in this respect.

However, if we want to summarise, we can still acknowledge the constructive role of medical imaging in the diagnosis of COVID-19.

### 4-3- Key aspects of automatic AI-based COVID-19 diagnosis

Due to the growing potential of AI-based approaches for medical diagnosis and interventions, using these approaches to distinguish COVID-19 from medical images has received much attention. With deep learning, the accuracy of the proposed methods has also increased dramatically. In this review study, 51 studies that used deep learning to diagnose COVID-19 were examined. To compare the studies conducted in this field, these studies have been summarised in terms of modality, methodology, accuracy and number of images used in Table 4. Summarised studies show an accuracy of over 80% in the diagnosis of COVID-19 based on deep learning methods. Therefore, this indicates the ability of deep learning methods in the analysis of medical images.

The main challenge in this section is the lack of big data for more accurate analysis. Although some studies have collected data, it is necessary to collect large dataset in this area due to using deep learning. However, the reviewed studies show the authors used approaches such as transfer learning and data augmentation to overcome this shortcoming. Fine-tuning of pre-trained neural networks for image classification can be an approach to network training with a small number of samples. Data augmentation using GANs or other image processing methods like image rotation and translation, lightening transformations, scaling and so on can also improve the learning and testing process by increasing the number of input images.

Although using deep neural networks has mostly been the basis for most articles in this field, hand-crafted feature extraction methods along with classification methods can also be evaluated for the COVID-19 diagnosis.

### 4-4- Outlook

In this study, two main issues related to COVID-19 were examined. First, the role of medical imaging in the COVID-19 diagnosis was investigated, and the details of the observed characteristics of this disease were listed based on different modalities. Second, AI-based automated methods for COVID-19 diagnosis in various images were reviewed to shed light on the importance of these techniques. This study could be useful for medical staff and technologists who want to get acquainted with the features of COVID-19 in medical images.

In future studies, it is possible to achieve more reliable results by collecting a much broader set of data from different medical centres and relying on approaches that simultaneously use multiple modalities and learning methods. Also, there is a need to develop an international protocol for using medical imaging to diagnose the severity of COVID-19 and its follow-up to control the destructive effects of medical imaging on patients, especially children and pregnant women.

**Table 1: Overview of studies on CT imaging that have five or more cases of COVID-19.**

| References | No. of cases | CT findings | Other findings |
|---|---|---|---|
| Ai et al. (2020) [98] | 1014 | GGOs, consolidation, reticulation/thickened interlobular septa and nodular lesions | Chest CT has a high sensitivity for the diagnosis of COVID-19. |
| Zhang et al. (2020) [40] | 645 | GGOs and consolidation | Combing clinical features and radiographic scores can effectively predict severe/critical types. |
| Ling et al. (2020) [36] | 295 | Four patients with COVID-19 infection showed no clinical symptoms or abnormal chest CT images | The clinical symptoms and radiological abnormalities are not the essential components of COVID-19 infection. |
| Yang et al. (2020) [99] | 273 | GGOs, consolidation and linear opacities, solid nodules, fibrous stripes, chronic inflammatory manifestation, chronic bronchitis, emphysema, pericardial effusion, pleural effusion, bullae of lung and obsolete tuberculosis | Age, Monocyte-lymphocyte ratio, homocysteine and period from onset to admission could predict imaging progression on chest CT from COVID-19 patients. |
| Colombi et al. (2020) [100] | 236 | Patchy GGO, diffuse GGO, GGO and consolidation, pleural effusion, mediastinal nodes enlargement, emphysema and pulmonary fibrosis | In patients with confirmed COVID-19 pneumonia, visual or software quantification the extent of CT lung abnormality were predictors of ICU admission or death. |
| Dai et al. (2020) [24] | 234 | Vascular enhancement sign, interlobular septal thickening, air bronchus sign, intralesional and/or perilesional bronchiectasis, pleural thickening, solid nodules, reticular/mosaic sign, interlobar fissure displacement, bronchial wall thickening, minor pleural effusion, pericardial effusion and mediastinal lymphadenopathy | Chest High-resolution CT provided the distribution, shape, attenuation and extent of lung lesions, and some typical CT signs of COVID-19 pneumonia. |
| Bai et al. (2020) [101] | 219 | GGO, fine reticular opacity and vascular thickening | High specificity but moderate sensitivity in distinguishing COVID-19 from viral pneumonia on chest CT. |
| Caruso et al. (2020) [102] | 158 | GGO, subsegmental vessel enlargement, consolidation, lymphadenopathy, bronchiectasis, air bronchogram, pulmonary nodules surrounded by GGO, interlobular septal thickening, halo sign, pericardial effusion, pleural effusion and bronchial wall thickening | Chest CT sensitivity was high (97%) but with lower specificity (56%). |
| Fan et al. (2020) [103] | 150 | Ground-glass nodules, patchy GGO, consolidation, cord-like lesions, thickening of the bronchovascular bundles, pleural thickening, crazy-paving sign, air bronchogram sign, pleural effusion and enlarged lymph nodes | The main manifestations of initial chest CT in COVID-19 is GGOs, commonly involving single site in patients < 35 years old and multiple sites and extensive area in patients > 60 years old. |
| Yang et al. (2020) [37] | 149 | GGO, mixed GGOs and consolidation, consolidation, air bronchogram, centrilobular nodules, tree-in-bud, reticular pattern, subpleural linear opacity, bronchial dilatation, cystic change, lymphadenopathy and pleural effusion | Some patients with COVID-19 can present with normal chest findings. |
| Li et al. (2020) [104] | 131 | GGOs, consolidation, nodule, interlobular septal thickening, vascular enlargement, air bronchogram, fibrosis, pleural thickening, hydrothorax and lymph node enlargement | The imaging pattern of multifocal peripheral ground glass or mixed consolidation is highly suspicious of COVID-19, that can quickly change over a short period. |
| Wu et al. (2020) [105] | 130 | GGO, GGO with consolidation, vascular thickening, pleural parallel sign, intralobular septal thickening, halo sign, reversed-halo sign, pleural effusion and pneumatocele | COVID-19 imaging characteristic mainly has subpleural, centrilobular and diffused distribution. The first two distributions can overlap or progress to diffused distribution. In the later period, it was mainly manifested as organising pneumonia and fibrosis. The most valuable characteristic is the pleural parallel sign. |
| Bernheim et al. (2020) [106] | 121 | GGOs, GGO with consolidation, consolidation, linear Opacities, rounded morphology of opacities, crazy paving pattern, reverse-halo sign, pleural effusion and underlying pulmonary emphysema | Recognising imaging patterns based on infection time course is paramount for helping to predict patient progression and potential complication development. |
| Zhang et al. (2020) [107] | 120 | GGOs, nodules, linear densities, consolidation, crazy paving, bronchiectasis, effusion, lymphadenopathy, air bronchograms, tree-in-bud sign and white lung | Using chest CT as the primary screening method in epidemic areas is recommended. |
| Zhao et al. (2020) [108] | 118 | GGO, consolidation, centrilobular nodules, architectural distortion, bronchial wall thickening, reticulation, subpleural bands, traction bronchiectasis, vascular enlargement, intrathoracic lymph node enlargement and pleural effusions | The follow-up CT changes during the treatment could help evaluate the treatment response of patients. |
| Wang et al. (2020) [34] | 114 | GGO, consolidation and pleural effusion | Spiral CT can make an early diagnosis and for evaluation of progression, with a diagnostic sensitivity and accuracy better than that of nucleic acid detection. |
| Han et al. (2020) [109] | 108 | GGO, consolidation, GGO with consolidation, vascular thickening, crazy paving pattern, air bronchogram sign and halo sign | |
| Zhao et al. (2020) [110] | 101 | GGOs, consolidation, mixed GGOs and consolidation, centrilobular nodules, architectural distortion, bronchial wall thickening, reticulation, subpleural bands, traction bronchiectasis, intrathoracic lymph node enlargement, vascular enlargement and pleural effusions | |
| Huang et al. (2020) [111] | 100 | GGO, consolidation, crazy-paving pattern, bronchiectasis, interlobular septal thickening and lymphadenopathy | The mechanism of CT features is explicable based on pathological findings. |
| Wang et al. (2020) [112] | 90 | GGO, consolidation, crazy-paving pattern and pleural effusion | The extent of CT abnormalities progressed rapidly after symptom onset, peaked during illness days 6-11, and followed by persistence of high levels. |
| Xu et al. (2020) [113] | 90 | GGO, consolidation, crazy-paving pattern, interlobular septal thickening, linear opacities combined, air bronchogram sign, adjacent pleura thickening, pleural effusion, pericardial effusion and lymphadenopathy | |
| Liang et al. (2020) [114] | 88 | GGO, consolidation, linear opacities, discrete pulmonary nodules and cavitation | |
| Li et al. (2020) [115] | 83 | GGO, linear opacities, consolidation, interlobular septal thickening, crazy-paving pattern, spider web sign, bronchial wall thickening, subpleural curvilinear line, nodule, reticulation, lymph node enlargement, pleural effusion and pericardial effusion | |
| Shi et al. (2020) [39] | 81 | Bilateral, subpleural, GGOs with air bronchograms, ill-defined margins, and a slight predominance in the right lower lobe, irregular interlobular septal thickening, crazy-paving pattern, thickening of the adjacent pleura, nodules, cystic changes, bronchiectasis, pleural effusion, lymphadenopathy, consolidation patterns and reticular patterns | CT findings vary depending on the time interval between the onset of symptoms and the CT performing. |
| Wu et al. (2020) [116] | 80 | GGO, consolidation, interlobular septal thickening, crazy-paving pattern, spider web sign, subpleural line, bronchial wall thickening, lymph node enlargement, pericardial effusion and pleural effusion | |
| Li et al. (2020) [35] | 78 | GGOs, mixed GGOs, consolidation, interlobular septal thickening, air bronchograms, fibrotic lesions and pleural effusion | No centrilobular nodules or lymphadenopathy. |
| Liu et al. (2020) [117] | 73 | Unique GGOs, multiple GGOs, paving stone sign, consolidation, bronchial wall thickening, pleural effusion and thickening of lung texture | The size and CT abnormalities are related to disease severity. |
| Zhu et al. (2020) [45] | 44 younger and 28 older | Pure ground-glass, GGO with consolidation, consolidation, reticular pattern or honeycombing, subpleural line, pleural thickening, pleural traction, pleural effusion, vacuolar sign, air bronchogram and vascular enlargement | Elderly and younger patients with COVID-19 have some similar CT features. However, older patients are more likely to have extensive lung lobe involvement, and subpleural line and pleural thickening. |
| Zhong et al. (2020) [118] | 67 | Solid plaque shadow, halo sign, fibrous strip shadow with ground-glass shadow and consolidation shadow | A solid shadow may predict severe and critical illness. |
| Pan et al. (2020) [28] | 63 | Patchy/punctate GGOs, GGOs, patchy consolidation, fibrous stripes and irregular solid nodules | |
| Zhou et al. (2020) [119] | 62 | GGO, consolidation, GGO with consolidation, nodule, rounded opacities, crazy-paving pattern, air bronchogram, halo sign, subpleural curvilinear line, thoracic lymphadenopathy, pleural effusion or thickening and pulmonary fibrosis | In patients with dyspno and respiratory distress, CT examination is very practical in the preclinical screening of patients with COVID-19. |
| Zhou et al. (2020) [120] | 62 | GGO, consolidation, GGO plus a reticular pattern, vacuolar sign, microvascular dilation sign, fibrotic streaks, subpleural line, subpleural transparent line, air bronchogram, bronchus distortion, thickening of pleura, pleural retraction sign and pleural effusion | GGO and a single lesion at the onset of COVID-19 pneumonia suggested that the disease was in its early phase. Pleural effusion might occur in the advanced phase. |
| Liu et al. (2020) [41] | 59 | Pure GGO, GGO with consolidation, GGO with reticulation, consolidation and pleural effusion | Atypical clinical findings of pregnant women with COVID-19 could increase the difficulty in initial identification. Consolidation was common in the pregnant groups. The chest CT imaging features of children with COVID-19 pneumonia were non-specific. At the same time, the exposure history and clinical symptoms could be more helpful for the screening. |
| Meng et al. (2020) [121] | 58 | GGO with peripheral distribution and unilateral location, fine reticulation, subpleural curvilinear line, halo sign, air bronchogram, vascular enlargement and consolidation | CT scan has great value in the highly suspicious, asymptomatic cases with negative nucleic acid testing. |
| Lomoro et al. (2020) [49] | 58 | GGO, GGO with consolidation, crazy-paving patterns, fibrous stripes, subpleural lines, architectural distortion, air bronchogram sign, perilesional vascular thickening, scattered nodules, enlarged mediastinal lymph nodes and pleural effusion | |
| Li and Xia(2020) [122] | 53 | GGOs and consolidation with or without vascular enlargement, interlobular septal thickening and air bronchogram sign | Low rate of misdiagnosis of COVID-19 in CT images. |
| Guan et al. (2020) [32] | 53 | GGO, crazy paving, consolidation, stripe, air bronchogram, pulmonary nodules and secondary tuberculosis | Identification of CT features of COVID-19 pneumonia provides timely diagnostic evidence. |
| Wang et al. (2020) [123] | 52 | GGOs, patchy consolidation and sub-consolidation, air bronchi sign, thickened leaflet interval and fibrous stripes | The chest CT images of patients with COVID-19 have specific characteristics with dynamic changes, which are of value for monitoring disease progress and clinical treatment. |
| Lyu et al. (2020) [124] | 51 | Consolidation, crazy-paving pattern and air bronchogram | Severity assessment of COVID-19 pneumonia based on chest CT would be feasible for critical cases. |
| Fang et al. (2020) [125] | 51 | GGOs, GGO with consolidation, consolidation and linear opacity | The sensitivity of CT for COVID-19 infection is 98% compared to RT-PCR sensitivity of 71%. |
| Xu et al. (2020) [126] | 50 | GGO, mixed GGOs and consolidation, consolidation, thickened intralobular septa, thickened interlobular septa, air bronchogram, pleural effusion and enlarged mediastinal nodes | Repeated CT scanning helps monitor disease progression and implement timely treatment. |
| Lei et al. (2020) [127] | 49 | GGOs, interstitial thickening, and consolidation, fibrosis, parenchymal band, traction bronchiectasis and irregular interfaces | |
| Yang et al. (2020) [128] | 44 | Pure GGOs, GGO with consolidation, GGO with interlobular septal thickening, consolidation, vessel expansion, air bronchogram, mediastinal lymphadenectasis and pleural effusion | The features of early-stage COVID-19 include GGO-based lesions with rare small size consolidation mainly distributed in the peripheral and posterior part of the lung. |
| Xiong et al. (2020) [29] | 42 | Single or multiple GGO, consolidation, interstitial thickening or reticulation, air bronchograms, pleural effusion and fibrous strips | |
| Long et al. (2020) [129] | 36 | GGOs, GGO with consolidation, lymphadenopathy and pleural effusion | Patients with typical CT findings but negative RRT-PCR results should be isolated. |
| Liu et al. (2020) [130] | 33 | Subpleural lesions, central lesions, ground-glass density shadow, consolidation, interstitial change and interlobular septal thickening | An important basis of CT images for early detection and disease monitoring. |
| Cheng et al. (2020) [131] | 11 COVID-19 and 22 non-COVID-19 | GGO, mixed GGO, consolidation, air bronchogram, centrilobular nodules, tree-in-bud sign, reticular pattern, subpleural linear opacity, bronchial dilatation and cystic change | findings of more extensive GGO than consolidation on chest CT scans obtained during the first week of illness were considered findings highly suspicious of COVID-19. |
| Yuan et al. (2020) [132] | 27 | GGO, consolidation, GGO and consolidation, air bronchogram, Nodular opacities and pleural effusion | A simple CT scoring method was capable of predicting mortality. |
| Dane et al. (2020) [133] | 23 | GGO, ground-glass nodule, solid nodule, consolidation, halo sign and interstitial thickening | |
| Wu et al. (2020) [31] | 23 | GGO, patchy, wedge-shaped ground-glass shadows, intralobular interstitial thickening with consolidation, fibrous stripes and concomitant hydropericardium and/or hydrothorax | Radiological findings and clinical characteristics in pregnant women with COVID-19 were similar to those of non-pregnant women with COVID-19. |
| Himoto et al. (2020) [134] | 21 | Bilateral GGO, peripheral-predominant lesions without airway abnormalities, mediastinal lymphadenopathy and pleural effusion | Important supplemental role of CT imaging to triage and detect patients suspected COVID-19 pneumonia, before getting the results of RT-PCR. |
| Chung et al. (2020) [135] | 21 | GGOs, GGO with consolidation, consolidation, rounded morphology, linear opacities and crazy-paving pattern | |
| Pan et al. (2020) [136] | 21 | GGOs, crazy-paving pattern, inter- and intralobular septal thickening and consolidation | Chest CT signs of improvement began at approximately 14 days after the onset of initial symptoms. |
| Xia et al. (2020) [42] | 20 | Consolidation with surrounding halo sign, GGOs, fine mesh shadow, tiny nodules, interlobular septal thickening, fibrosis lesions, air bronchogram signs and pleural thickening | Procalcitonin elevation and consolidation with surrounding halo signs were frequent in paediatric patients. |
| Zhu et al. (2020) [137] | 7 patients with Heart failure and 12 with COVID-19 | GGO and thickening of the interlobular septum in both group. In heart failure group, the ratio of the expansion of small pulmonary veins was also higher. | There are significant differences in chest CT features, such as enlargement of pulmonary veins, lesions distribution and morphology between heart failure and COVID-19. |
| Han et al. (2020) [33] | 17 | GGO, GGO with interlobular septal thickening, GGO with irregular linear opacities, consolidation, presence of nodule, enlarged pulmonary vessels, bronchiolar dilatation, crazy paving, air bronchogram, thickening of the adjacent pleura, interleaf fissure displacement, evidence of pulmonary fibrosis and pleural effusion | There is a synchronised improvement in both clinical and radiologic features in the 4th week. |
| Feng et al. (2020) [46] | 15 | Small nodular GGOs and speckled GGOs | Dynamic reexamination of chest CT and nucleic acid are essential in children. |
| Lei et al. (2020) [138] | 14 | Presence of nodular, GGO, bronchovascular enlarged, irregular linear appearances, consolidation pulmonary opacity and pleural effusion | |
| Zhu et al. (2020) [38] | 14 | GGOs, mixed GGO and consolidation, reticulation, crazy paving, cavitation and bronchiectasis | There is a need to develop a new detection technique. |
| Chate et al. (2020) [139] | 12 | GGOs, crazy-paving pattern, alveolar consolidation, reversed-halo sign and pleural effusion | |
| Agostini et al. (2020) [30] | 10 | GGOs, GGO with consolidation, linear opacities, rounded opacities, crazy-paving pattern, reverse-halo sign, bronchial wall thickening and bronchiectasis | Ultra-low-dose, dual-source, fast CT protocol provides highly diagnostic images for COVID-19 with potential for reduction in dose and motion artefacts. |
| Zhou et al. (2020) [43] | 9 | Nodular lesions, patchy lesions, GGO with consolidation and halo sign | Infants and young children with COVID-19 have mild clinical symptoms and imaging findings not as typical as those of adults. |
| Yoon et al. (2020) [51] | 9 | Pure GGO, mixed GGO and consolidation, consolidation, crazy-paving appearance and air bronchogram | |
| Iwasawa et al. (2020) [26] | 6 | GGOs, consolidation, linear opacities, reticulation and crazy-paving pattern | U-HRCT can evaluate not only the distribution and hallmarks of COVID-19 pneumonia but also visualise local lung volume loss. |
| Gao and Zhang (2020) [25] | 6 | GGOs, nodule, halo sign, thickened lobular septum, thickened bronchial wall, tree-in-bud sign, crazy-paving sign, proliferation and calcification | The imaging manifestations of early-stage COVID-19 are relatively mild, and the imaging findings of some patients are not typical, which can easily lead to missed diagnoses. |
| Zhu et al. (2020) [140] | 6 | GGO, GGO with consolidation, consolidation, reticulation, crazy paving and bronchiectasis | In the early-stage of the disease, the lesion can manifest as round nodular-like GGO in the central area of the lung lobe. The follow-up CT images showed the lesions are migratory manifested as the absorption of the primary lesions and the emergence of new lesions. |
| Li et al. (2020) [44] | 5 | Patchy GGOs | Similar but more modest lung abnormalities at CT of children compared to adults |
| Liu et al. (2020) [27] | 5 | GGOs with consolidation | The paediatric patients generally have milder CT findings than adults. |
| Lu and Pu (2020) [141] | 5 | Crazy-paving pattern, GGOs, septal line thickening, consolidation and thickened interlobular septa | |
| Xie et al. (2020) [142] | 5 | Multifocal GGO, parenchyma consolidation, mixed GGO and mixed consolidation | |

## Table 2: Overview of case reports and letters to the editors on CT imaging that have less than five cases of COVID-19.

| Reference | Remarks |
|---|---|
| McGinnis et al. (2020) [143] | Asymptomatic COVID-19 was detected using CT imaging in a patient with recurrent non-small cell lung cancer. |
| Yan et al. (2020) [144] | Chest CT findings are important when there is a false-positive results for COVID-19. |
| Qi et al. (2020) [145] | CT imaging can play an important role in managing patients of COVID-19 for diagnosis and monitoring. |
| Zhang et al. (2020) [146] | CT imaging can be helpful for early detection of COVID-19 based on CT findings. |
| Tenda et al. (2020) [147] | Three patients with mild to moderate symptoms have been considered. They highly suggest the use of non-contrast chest CT for COVID-19 diagnosis in patients with moderate symptoms. |
| Erturk (2020) [148] | CT may help diagnose but not screening highly suspected cases. |
| Xu et al. (2020) [149] | Six patients from an extended family with COVID-19 have been investigated. It should be avoided to rely on CT for clinical diagnosis. |
| Liu et al. (2020) [150] | Chest CT has an indispensable role in early detection and diagnosis of COVID-19 infection, however, further investigation is needed. |
| Li et al. (2020) [151] | Repeated CT scanning could facilitate monitoring disease progression and implementing proper treatment. |
| Ufuk (2020) [152] | A patient with peripheral, multilobar areas of GGO sign in chest CT images and positive for COVID-19 has been presented. |
| Hamer et al. (2020) [153] | A positive case of COVID-19 and also a review of some studies in this field have been presented. CT morphology can be a support for COVID-19 diagnosis. |
| Kang et al. (2020) [154] | A low-dose scanning protocol has been presented that reduces the patient’s dose to 1/8 to 1/9 of the standard dose without significant sacrifice of signal-to-noise (SNR) or contrast-to-noise (CNR) ratios. |
| Wang et al. (2020) [48] | The role of CT in the diagnosis of COVID-19 is not clear, so it is better to use alternative modalities such as Ultrasound or CXR due to lower radiation, especially for children. |
| Zou and Zhu (2020) [58] | A case with GGOs with areas of focal consolidation primarily in the right upper lobe and a focal opacity in the left upper and right middle lobes has been reported. |
| Dai et al. (2020) [155] | For patients with fever as the first symptom and with a history of exposure to COVID-19, chest CT examination should be performed soon. |
| Lin et al. (2020) [156] | Observe changes in CT images during the disease. |
| Lee et al. (2020) [157] | More research is needed into the correlation of CT findings with clinical severity and progression of COVID-19. |
| Kim (2020) [158] | Role of radiologists includes not only early detection of lung abnormality, but also suggestion of disease severity, potential progression to acute respiratory distress syndrome, and possible bacterial co-infection in hospitalised patients. |
| Zhang et al. (2020) [159] | One case of COVID-19 pneumonia showed multiple subpleural GGOs in bilateral lung, rapid progression, and it also accompanied nodular GGOs on chest CT. |
| Singh and Fratesi (2020) [160] | CT may expedite care in symptomatic patients with a negative or pending swab, and in those with worsening respiratory status or developing complications such as empyema or acute respiratory distress syndrome. |
| Tsou et al. (2020) [161] | In Singapore, the consensus of the infectious diseases experts is to rely on reverse transcriptase polymerase chain reaction (RT-PCR) for diagnosis rather than to use CT. |
| Vu et al. (2020) [162] | Patients not initially suspected of COVID-19 infection can be quarantined earlier to limit exposure to others using CT imaging features of COVID-19. |
| Çinkooğlu et al. (2020) [163] | Imaging plays a critical role in initial diagnosis and in assessment of disease severity and progression. |
| Asadollahi-Amin et al. (2020) [164] | A patient found to be positive COVID-19 after a CT scan performed for an unrelated condition revealed a lesion in the lung field compatible with COVID-19 infection. |
| Chen et al. (2020) [165] | Chest CT scan is the primary diagnostic approach for COVID-19 and its feature includes multiple, bilateral, patchy consolidation and GGO with subpleural distribution. |
| Hu et al. (2020) [166] | Two cases both demonstrated symptom relief but progression on CT, which indicates that clinical symptoms and imaging findings are inconsistent in early-stage of COVID-19 pneumonia. |
| Lim et al. (2020) [167] | This case series of three COVID-19 pneumonia patients highlights the variable chest CT features during the acute and convalescent phases. Chest CT is a highly sensitive tool for the delineation of the extent of lung disease. However, its use as a first-line diagnostic modality to replace RT-PCR is not guaranteed. |
| Ostad et al. (2020) [168] | Radiologists should know the possibility of artefacts when reporting the axial CT images with limited involvement, especially those cases with focal basal GGO, where linear atelectasis is also common. In such cases correlation with reformatted planes and utilising thin-section reconstructions are recommended to avoid misinterpretation. |
| Qanadli et al. (2020) [169] | Vascular findings convey both diagnostic and prognostic information and might contribute to disease diagnosis and patient management. The vascular congestion sign may help distinguish COVID-19 from community-acquired pneumonia. |
| Mungmungpuntipantip and Wiwanitkit (2020) [170] | Chest CT cannot discriminate early COVID-19 from other diseases. |
| Danrad et al. (2020) [171] | Early and advanced stage CT finding from patients with documented COVID-19 admitted to University medical center in New Orleans, Louisiana have been described. |
| Joob and Wiwanitkit (2020) [172] | If patients have underlying lung disease such as tuberculosis, atypical chest CT findings might be seen. Practitioners have to recognise the broad spectrum of possible CT findings in patients with COVID-19. |
| Li et al. (2020) [173] | Characteristic imaging changes were found with GGO, consolidation and septal thickening mainly distributed in peripheral and posterior area by thoracic CT scan in the three patients. |
| Guan et al. (2020) [174] | A dynamic chest CT scan plays a significant role in the diagnosis and prognosis of COVID-19. |
| Li et al. (2020) [175] | CT plays a vital role in the diagnosis, staging, and monitoring of patients with COVID-19 pneumonia. |
| Lei et al. (2020) [176] | Knowing the corresponding CT feature of COVID-19 pneumonia at different stages, which could be helpful to precisely diagnose and understand CT characteristics of COVID-19. |
| Gross et al. (2020) [177] | CT may be a useful tool to evaluate the extent of the disease in severe cases, provide prognostic information and guide future treatment options. |
| Shi et al. (2020) [178] | COVID-19 pneumonia may present with atypical manifestations, such as haemoptysis and focal GGO with non-peripheral distribution, on initial CT scans. |
| Qu et al. (2020) [179] | This study is a report of manifestations of COVID-19 in a patient with lung adenocarcinoma. |
| An et al. (2020) [180] | Chest CT offers fast and convenient evaluation of patients with suspected COVID-19 pneumonia. |
| Wei et al. (2020) [181] | CT showed rapidly progressing peripheral consolidations and GGOs in both lungs of a 40-year-old female patient with COVID-19 pneumonia. After treatment, the lesions were almost absorbed leaving the fibrous lesions. |
| Fang et al. (2020) [182] | Under the circumstances, computed tomography imaging is not only useful for the detection, location of lesions but also helpful in evaluating the dynamic changes of patients with COVID-19. CT imaging can play a determinant role in clinical decision-making. |
| Duan and Qin (2020) [183] | At seven days, chest CT showed decreasing GGOs in a 46-year-old woman. At day 13 after admission, the GGOs in the right lung had resolved; the left GGOs showed partial resolution. |
| Shi et al. (2020) [184] | This study uses imaging data for patient's improvement monitoring in a case with COVID-19. |
| Fang et al. (2020) [185] | The authors report two cases of COVID-19 using CT imaging data. |
| Kanne (2020) [186] | In the correct clinical setting, bilateral GGOs or consolidation at chest imaging should prompt the radiologist to suggest COVID-19 as a possible diagnosis. A normal chest CT scan does not exclude the diagnosis of COVID-19 infection. |
| Adair and Ledermann (2020) [187] | This case report discusses the imaging findings of one of the first cases in the mid-western United States. |
| Burhan et al. (2020) [188] | The result may suggest that in an area with high number of COVID-19 case, CT Scan might be a better diagnostic tool compared to RT-PCR in diagnosing COVID-19. |
| Feng et al. (2020) [189] | It is challenging to distinguish COVID-19 pneumonia from other viral pneumonia on CT findings alone; however, the authors emphasise the utility of chest CT to detect early change of COVID-19 in cases which RT-PCR tests show negative results. |
| Hao and Li (2020) [190] | If patients have clinical symptoms, epidemiological characteristics, and chest CT imaging characteristics of viral pneumonia compatible with COVID-19 infection, we need to carefully consider the isolation and treatment of these patients even if the RT-PCR test is negative. |
| Yang and Yan (2020) [191] | A patient with RT-PCR-confirmed COVID-19 infection may have normal chest CT at admission. |
| Lei et al. (2020) [192] | The bilateralism of the peripheral lung opacities, without subpleural sparing, are common CT findings of COVID-19 pneumonia. |

## Table 3: Overview of studies on CXR imaging and related findings.

| References | No. of cases | Findings |
|---|---|---|
| Jacobi et al. (2020) [47] | - | Irregular, patchy, hazy, reticular and widespread ground-glass opacities. |
| Lomoro et al. (2020) [49] | 58 | Consolidation and hazy increased opacity. |
| Wong et al. (2020) [50] | 64 | GGOs, consolidation and pleural effusion. |
| Zhang et al. (2020) [40] | 645 | GGOs and consolidation. |
| Yoon et al. (2020) [51] | 9 | Parenchymal abnormalities, consolidation, GGOs, single nodular opacity and patchy opacities. |
| Wang et al. (2020) [48] | - | It is better to use CXR due to lower radiation, especially for children. |
| Shi et al. (2020) [184] | 1 | This study uses imaging data for patient's improvement monitoring in a case with COVID-19. |

## Table 4: Overview of deep learning approaches for automated COVID-19 diagnosis.

| Reference | Task | Modality | Method | Total No. Of Images | No. Of Images From COVID-19 Cases | Accuracy (%) | Remarks |
|---|---|---|---|---|---|---|---|
| Wang and Wong (2020) [93] | Automatic COVID-19 diagnosis | CXR | CNN | 13,975 | 358 | 93.3 | COVID-Net has been proposed. |
| Narin et al. (2020) [84] | Automatic COVID-19 diagnosis | CXR | CNN | 100 | 50 | 98 | The pre-trained ResNet50 model provides the highest classification performance. |
| Hemdan et al. (2020) [85] | Automatic COVID-19 diagnosis | CXR | CNN | 75 | 25 | 90 | The VGG19 and DenseNet201 models showed a good and similar performance. |
| Ghoshal and Tucker (2020) [86] | Estimating uncertainty and interpretability in deep learning for COVID-19 diagnosis | CXR | BCNN | 5,941 | 68 | 89 | Experiment has shown a strong correlation between model uncertainty and accuracy of prediction. |
| Apostolopoulos and Mpesiana (2020) [193] | Automatic COVID-19 diagnosis | CXR | CNN | 1,442 | 224 | 96.78 | The MobileNet v2 effectively distinguished the Covid-19 cases from viral and bacterial pneumonia cases. |
| Zhang et al. (2020) [194] | Anomaly detection for fast and reliable COVID-19 screening | CXR | CNN | 1,531 | 100 | 96 | The model comprises a backbone network, a classification head, and an anomaly detection head. |
| Farooq and Hafeez (2020) [89] | Automatic COVID-19 diagnosis | CXR | CNN | 5,941 | 68 | 96.23 | COVID-ResNet for classification of COVID-19 and three other infection types has been proposed. |
| Sethy and Behera (2020) [195] | Automatic COVID-19 diagnosis | CXR | CNN | 50 | 25 | 95.38 | ResNet50 plus SVM have been used for COVID-19 diagnosis. |
| Apostolopoulos et al. (2020) [90] | Automatic classification of pulmonary diseases | CXR | CNN | 3,905 | 455 | 99.18 | Mobile Net has been used for transfer learning. |
| Abbas et al. (2020) [196] | Automatic COVID-19 diagnosis | CXR | CNN | 196 | 105 | 95.12 | A deep CNN, called Decompose, Transfer, and Compose (DeTraC) has been validated. |
| Afshar et al. (2020) [197] | Automatic COVID-19 diagnosis | CXR | CNN | 13,975 | 358 | 95.7 | COVID-CAPS including several Capsule and convolutional layers has been proposed. |
| Chowdhury et al. (2020) [198] | Automatic COVID-19 diagnosis | CXR | CNN | 2,876 | 190 | 98.3 | SqueezeNet outperforms AlexNet, ResNet18 and DenseNet201. |
| Li and Zhu (2020) [199] | Automatic COVID-19 diagnosis | CXR | CNN | 537 | 179 | 89.7 | COVID-MobileXpert: a lightweight deep neural network based mobile app that can use noisy snapshots of CXR for point-of-care COVID-19 screening has been presented. |
| Karim et al. (2020) [200] | Automatic COVID-19 diagnosis | CXR | CNN | 16,995 | 259 | - | Evaluation results based on hold-out data show a positive predictive value (PPV) of 89.61% and recall of 83%. |
| Oh et al. (2020) [94] | Automatic COVID-19 diagnosis | CXR | CNN | 15,043 | 180 | 91.9 | A patch-based deep neural network architecture that can be stably trained with small data set has been proposed. |
| Hall et al. (2020) [201] | Automatic COVID-19 diagnosis | CXR | CNN | 455 | 135 | 91.24 | Resnet50 was tuned on 102 COVID-19 cases and 102 other pneumonia cases in a 10-fold cross validation. |
| Rajaraman et al. (2020) [202] | Automatic COVID-19 diagnosis | CXR | CNN | 16,700 | 313 | 99.01 | The best performing models are iteratively pruned to identify optimal number of neurons in the convolutional layers to reduce complexity and improve memory efficiency. |
| Luz et al. (2020) [203] | Automatic COVID-19 diagnosis | CXR | CNN | 13,800 | 183 | 93.9 | The proposed model has about 30 times parameters fewer than the baseline literature model, 28 and 5 times parameters fewer than the popular VGG16 and ResNet50 architectures, respectively. |
| Tartaglione et al. (2020) [204] | Automatic COVID-19 diagnosis | CXR | CNN | 584 | 405 | 95 | Possible obstacles in successfully training a deep model have been highlighted. |
| Ezzat and Ella (2020) [205] | Automatic COVID-19 diagnosis | CXR | CNN | 5,961 | 126 | 98 | The proposed GSA-DenseNet121-COVID19 approach is consists of data preparation, hyper-parameters selection, learning and performance measurement stage. |
| Hammoudi et al. (2020) [206] | Automatic COVID-19 diagnosis | CXR | CNN | 5,863 | - | 95.72 | The DenseNet169 architecture has reached the best performance. |
| Kumar et al. (2020) [207] | Automatic COVID-19 diagnosis | CXR | CNN | 8,588 | 62 | 97.7 | The machine learning-based classification of the extracted deep feature using ResNet152 has been reported with the highest accuracy for XGBoost predictive classifier. |
| Khan et al. (2020) [208] | Automatic COVID-19 diagnosis | CXR | CNN | 1,300 | 284 | 89.5 | CoroNet, a deep CNN based model, has been proposed. |
| Khalifa et al. (2020) [87] | Automatic COVID-19 diagnosis | CXR | CNN and GAN | 5,863 | - | 99 | Resnet18 is the most appropriate deep transfer model according to testing accuracy measurement with using GAN as an image augmentation method. |
| Santosh et al. (2020) [209] | Automatic COVID-19 diagnosis | CXR | CNN | 6,756 | 73 | 99.96 | The Truncated Inception Net deep learning model has been proposed. |
| Khobahi et al. (2020) [210] | Screening COVID-19 at early and intermediate stages | CXR | CNN and AE | 18,529 | 99 | 93.5 | A semi-supervised methodology and two auto-encoders for automatic segmentation of the infected regions have been presented. |
| Rahimzadeh and Attar (2020) [211] | Automatic COVID-19 diagnosis | CXR | CNN | 15,085 | 180 | 99.56 | A concatenated neural network based on Xception and ResNet50V2 networks has been proposed. |
| Loey et al. (2020) [212] | Automatic COVID-19 diagnosis | CXR | CNN and GAN | 306 | 69 | 100 | A GAN for data augmentation with deep transfer learning has been presented. |
| Pereira et al. (2020) [213] | Automatic COVID-19 diagnosis | CXR | CNN | 1,144 | 90 | - | A macro-avg F1-Score of 0.65 using a multi-class approach and an F1-Score of 0.89 for the COVID-19 identification in the hierarchical classification scenario have been achieved. |
| Gozes et al. (2020) [214] | Automatic COVID-19 diagnosis | CT | CNN | 270 | 120 | - | Classification result for Coronavirus vs Non-coronavirus cases per thoracic CT studies is 0.996 AUC. |
| Shan et al. (2020) [215] | Automatic segmentation and quantification of infection regions | CT | CNN | 549 | 549 | 91.6 | Dice similarity coefficients of 91.6%±10.0% between automatic and manual segmentations have been reported. |
| Butt et al. (2020) [216] | Automatic COVID-19 diagnosis | CT | CNN | 618 | 219 | - | An AUC of 0.996 (95%CI: 0.989–1.00) for Coronavirus vs Non-coronavirus cases per thoracic CT studies has been achieved. |
| Wang et al. (2020) [217] | Automatic COVID-19 diagnosis | CT | CNN | 453 | 195 | 82.9 | Feature extraction has been done using transfer learning. |
| Li et al. (2020) [218] | Automatic COVID-19 diagnosis | CT | CNN | 4,356 | 1,296 | - | An AUC of 0.96 for detecting COVID-19 has been achieved. |
| Huang et al. (2020) [219] | Evaluation of lung burden changes in patients with COVID-19 | CT | CNN | 126 | 126 | - | A commercially available deep-learning-based tool has been used. |
| Zheng et al. (2020) [91] | Automatic COVID-19 diagnosis | CT | CNN | 630 | Not mentioned | 90.1 | A pre-trained U-Net for lung segmentation and a 3D CNN architecture (DeCoVNet) have been used. |
| Song et al. (2020) [220] | Automatic COVID-19 diagnosis | CT | CNN | 1,990 | 777 | 94 | A Details Relation Extraction neural network (DRE-Net) to extract the top-K details in the CT images has been designed. |
| Gozes et al. (2020) [221] | Detection, localising and quantifying the severity of COVID-19 manifestation | CT | CNN | 270 | 120 | - | An AUC of 0.948 has been achieved. |
| Chen et al. (2020) [222] | Automatic COVID-19 diagnosis | CT | CNN | 35,355 | 20,886 | 95.24 | U-NET++ has been used for retrospective and prospective COVID-19 dataset evaluation. |
| Chaganti et al. (2020) [223] | Automatic COVID-19 diagnosis | CT | CNN | 9,223 | 148 | - | Pearson Correlation Coefficient between method prediction and ground truth is 0.95 (Percentage of Opacity), 0.98 (Percentage of High Opacity), 0.96 (Lung Severity Score), 0.96 (Lung High Opacity Score). |
| He et al. (2020) [224] | Automatic COVID-19 diagnosis | CT | CNN | 746 | 349 | 86 | The approach achieves an F1 of 0.85 and an AUC of 0.94 in a small data set using a self-supervised transfer learning approach. |
| Hu et al. (2020) [92] | Automatic COVID-19 diagnosis | CT | CNN | 450 | 150 | 96.2 | A weakly-supervised deep learning framework for fast and fully-automated detection and classification of COVID-19 has been presented. |
| Fu et al. (2020) [225] | Automatic COVID-19 diagnosis | CT | CNN | 89,628 | 14,944 | 98.8 | A pre-trained ResNet-50 has been used to recognise COVID-19, non-COVID-19 viral pneumonia, bacterial pneumonia, pulmonary tuberculosis and healthy lung. |
| Loey et al. (2020) [88] | Automatic COVID-19 diagnosis | CT | CNN and CGAN | 742 | 345 | 82.91 | Data augmentations along with CGAN improve the performance of classification in AlexNet, VGGNet16, VGGNet19, GoogleNet, and ResNet50 deep transfer models. |
| Chen et al. (2020) [226] | Automated multi-class segmentation of COVID-19 chest CT images | CT | CNN | 110 | 110 | 89 | A residual attention U-Net for the lung CT image segmentation has been proposed. |
| Zhou et al. (2020) [227] | Automatic COVID-19 CT segmentation | CT | CNN | 473 | 473 | - | A U-Net based segmentation network using attention mechanism has been presented. The obtained Dice Score, Sensitivity and Specificity are 83.1%, 86.7% and 99.3%, respectively. |
| Mobiny et al. (2020) [228] | Automatic COVID-19 diagnosis | CT | CNN and CGAN | 746 | 349 | 87.6 | Detail-oriented capsule networks (DECAPS) for COVID-19 CT classification and CGAN as a data augmentation technique have been presented. |
| Wu et al. (2020) [229] | Classification and Segmentation for COVID-19 diagnosis | CT | CNN | 144,167 | 68,626 | - | A Joint Classification and Segmentation (JCS) system obtains an average sensitivity of 95.0% and a specificity of 93.0% on the classification test set, and 78.3% Dice score on the segmentation test set. |
| Maghdid et al. (2020) [230] | Automatic COVID-19 diagnosis | CXR and CT | CNN | CXR: 170<br>CT: 361 | CXR: 85<br>CT: 203 | 98 | The utilised models can provide accuracy up to 98% via pre-trained AlexNet and 94.1% accuracy by using the modified CNN. |
| Alom et al. (2020) [231] | Automatic COVID-19 diagnosis | CXR and CT | CNN | Not clear | Not clear | CXR: 84.67<br>CT: 98.78 | An improved Inception recurrent residual neural network (IRRCNN) and NABLA-3 network models were applied. |
| Razzak et al. (2020) [232] | Automatic COVID-19 diagnosis | CXR and CT | CNN | 800 | 200 | 98.75 | The results have been obtained using transfer learning. |

**References:**

1. Organization, W.H. *WHO Coronavirus Disease (COVID-19) Dashboard*. Available from: https://covid19.who.int/.
2. Organization, W.H. *Coronavirus disease (COVID-19) advice for the public*. Available from: https://www.who.int/emergencies/diseases/novel-coronavirus-2019/advice-for-public.
3. Rodrigues, J., et al., *An update on COVID-19 for the radiologist-A British society of Thoracic Imaging statement.* Clinical radiology, 2020. **75**(5): p. 323-325.
4. Nair, A., et al., *A British Society of Thoracic Imaging statement: considerations in designing local imaging diagnostic algorithms for the COVID-19 pandemic.* Clinical Radiology, 2020. **75**(5): p. 329-334.
5. Huang, Y., et al., *CT screening for early diagnosis of SARS-CoV-2 infection.* The Lancet Infectious Diseases, 2020.
6. Raptis, C.A., et al., *Chest CT and Coronavirus disease (COVID-19): a critical review of the literature to date.* American Journal of Roentgenology, 2020: p. 1-4.
7. Chua, F., et al., *The role of CT in case ascertainment and management of COVID-19 pneumonia in the UK: insights from high-incidence regions.* The Lancet Respiratory Medicine, 2020.
8. Duan, Y.-n., et al., *CT features of novel coronavirus pneumonia (COVID-19) in children.* European Radiology, 2020: p. 1-7.
9. Fan, L., et al., *Progress and prospect on imaging diagnosis of COVID-19.* Chinese Journal of Academic Radiology, 2020: p. 1-10.
10. Li, B., et al., *Diagnostic value and key features of computed tomography in Coronavirus Disease 2019.* Emerging Microbes & Infections, 2020(just-accepted): p. 1-14.
11. Majidi, H. and F. Niksolat, *Chest CT in patients suspected of COVID-19 infection: A reliable alternative for RT-PCR.* The American Journal of Emergency Medicine, 2020.
12. Nasir, M.U., et al., *The Role of emergency radiology in COVID-19: from preparedness to diagnosis.* Canadian Association of Radiologists Journal, 2020: p. 0846537120916419.
13. Plesner, L.L., et al., *Diagnostic imaging findings in COVID-19.* Ugeskrift for Laeger, 2020. **182**(15).
14. Rubin, G.D., et al., *The role of chest imaging in patient management during the COVID-19 pandemic: a multinational consensus statement from the Fleischner Society.* Chest, 2020.
15. Sahu, K.K., A. Lal, and A.K. Mishra, *An update on CT chest findings in coronavirus disease-19 (COVID-19).* Heart & Lung, 2020.
16. Sun, Z., *Diagnostic Value of Chest CT in Coronavirus Disease 2019 (COVID-19).* Current medical imaging, 2020.
17. Yang, W., et al., *The role of imaging in 2019 novel coronavirus pneumonia (COVID-19).* European Radiology, 2020: p. 1-9.
18. Ye, Z., et al., *Chest CT manifestations of new coronavirus disease 2019 (COVID-19): a pictorial review.* European radiology, 2020: p. 1-9.
19. Zhu, J., et al., *CT imaging features of 4,121 patients with COVID-19: a meta-analysis.* Journal of Medical Virology, 2020.
20. Li, M., *Chest CT features and their role in COVID-19.* Radiology of Infectious Diseases, 2020.
21. Hani, C., et al., *COVID-19 pneumonia: A review of typical CT findings and differential diagnosis.* Diagnostic and interventional imaging, 2020.
22. Bao, C., et al., *Coronavirus Disease 2019 (COVID-19) CT Findings: A Systematic Review and Meta-analysis.* Journal of the American College of Radiology, 2020.
23. Salehi, S., et al., *Coronavirus disease 2019 (COVID-19): a systematic review of imaging findings in 919 patients.* American Journal of Roentgenology, 2020: p. 1-7.
24. Dai, H., et al., *High-resolution chest CT features and clinical characteristics of patients infected with Covid-19 in Jiangsu, China.* International Journal of Infectious Diseases, 2020.
25. Gao, L. and J. Zhang, *Pulmonary High-Resolution Computed Tomography (HRCT) Findings of Patients with Early-Stage Coronavirus Disease 2019 (COVID-19) in Hangzhou, China.* Medical

Science Monitor: International Medical Journal of Experimental and Clinical Research, 2020. **26**: p. e923885-1.
26. Iwasawa, T., et al., *Ultra-high-resolution computed tomography can demonstrate alveolar collapse in novel coronavirus (COVID-19) pneumonia.* Japanese Journal of Radiology, 2020: p. 1.
27. Liu, M., Z. Song, and K. Xiao, *High-Resolution Computed Tomography Manifestations of 5 Pediatric Patients With 2019 Novel Coronavirus.* Journal of Computer Assisted Tomography, 2020.
28. Pan, Y., et al., *Initial CT findings and temporal changes in patients with the novel coronavirus pneumonia (2019-nCoV): a study of 63 patients in Wuhan, China.* European radiology, 2020: p. 1-4.
29. Xiong, Y., et al., *Clinical and high-resolution CT features of the COVID-19 infection: comparison of the initial and follow-up changes.* Investigative radiology, 2020.
30. Agostini, A., et al., *Proposal of a low-dose, long-pitch, dual-source chest CT protocol on third-generation dual-source CT using a tin filter for spectral shaping at 100 kVp for CoronaVirus Disease 2019 (COVID-19) patients: a feasibility study.* Radiol Med, 2020. **125**(4): p. 365-373.
31. Wu, X., et al., *Radiological findings and clinical characteristics of pregnant women with COVID-19 pneumonia.* International Journal of Gynecology & Obstetrics, 2020.
32. Guan, C.S., et al., *Imaging features of coronavirus disease 2019 (COVID-19): evaluation on thin-section CT.* Academic radiology, 2020.
33. Han, X., et al., *Novel Coronavirus Pneumonia (COVID-19) Progression Course in 17 Discharged Patients: Comparison of Clinical and Thin-Section CT Features During Recovery.* Clinical Infectious Diseases, 2020.
34. Wang, K., et al., *Imaging manifestations and diagnostic value of chest CT of coronavirus disease 2019 (COVID-19) in the Xiaogan area.* Clinical radiology, 2020.
35. Li, K., et al., *CT image visual quantitative evaluation and clinical classification of coronavirus disease (COVID-19).* European Radiology, 2020: p. 1-10.
36. Ling, Z., et al., *Asymptomatic SARS-CoV-2 infected patients with persistent negative CT findings.* European journal of radiology, 2020. **126**.
37. Yang, W., et al., *Clinical characteristics and imaging manifestations of the 2019 novel coronavirus disease (COVID-19): A multi-center study in Wenzhou city, Zhejiang, China.* Journal of Infection, 2020.
38. Zhu, Y., et al., *Clinical and CT imaging features of 2019 novel coronavirus disease (COVID-19).* Journal of Infection, 2020.
39. Shi, H., et al., *Radiological findings from 81 patients with COVID-19 pneumonia in Wuhan, China: a descriptive study.* The Lancet Infectious Diseases, 2020.
40. Zhang, X., et al., *Epidemiological, clinical characteristics of cases of SARS-CoV-2 infection with abnormal imaging findings.* International Journal of Infectious Diseases, 2020.
41. Liu, H., et al., *Clinical and CT imaging features of the COVID-19 pneumonia: Focus on pregnant women and children.* Journal of infection, 2020.
42. Xia, W., et al., *Clinical and CT features in pediatric patients with COVID-19 infection: Different points from adults.* Pediatric pulmonology, 2020. **55**(5): p. 1169-1174.
43. Zhou, Y., et al., *Clinical features and chest CT findings of coronavirus disease 2019 in infants and young children.* Zhongguo Dang dai er ke za zhi= Chinese Journal of Contemporary Pediatrics, 2020. **22**(3): p. 215-220.
44. Li, W., et al., *Chest computed tomography in children with COVID-19 respiratory infection.* Pediatric radiology, 2020: p. 1-4.
45. Zhu, T., et al., *A Comparative Study of Chest Computed Tomography Features in Young and Older Adults With Corona Virus Disease (COVID-19).* Journal of Thoracic Imaging, 2020.
46. Feng, K., et al., *Analysis of CT features of 15 children with 2019 novel coronavirus infection.* Zhonghua er ke za zhi= Chinese journal of pediatrics, 2020. **58**: p. E007-E007.

47. Jacobi, A., et al., *Portable chest X-ray in coronavirus disease-19 (COVID-19): A pictorial review.* Clinical Imaging, 2020.
48. Wang, Y.X.J., et al., *The role of CT for Covid-19 patient's management remains poorly defined.* Annals of Translational Medicine, 2020. **8**(4).
49. Lomoro, P., et al., *COVID-19 pneumonia manifestations at the admission on chest ultrasound, radiographs, and CT: single-center study and comprehensive radiologic literature review.* European Journal of Radiology Open, 2020: p. 100231.
50. Wong, H.Y.F., et al., *Frequency and distribution of chest radiographic findings in COVID-19 positive patients.* Radiology, 2020: p. 201160.
51. Yoon, S.H., et al., *Chest radiographic and CT findings of the 2019 novel coronavirus disease (COVID-19): analysis of nine patients treated in Korea.* Korean journal of radiology, 2020. **21**(4): p. 494-500.
52. Lu, W., et al., *A Clinical Study of Noninvasive Assessment of Lung Lesions in Patients with Coronavirus Disease-19 (COVID-19) by Bedside Ultrasound.* Ultraschall in der Medizin-European Journal of Ultrasound, 2020.
53. Vetrugno, L., et al., *Our Italian Experience Using Lung Ultrasound for Identification, Grading and Serial Follow-up of Severity of Lung Involvement for Management of Patients with COVID-19.* Echocardiography, 2020.
54. Kalafat, E., et al., *Lung ultrasound and computed tomographic findings in pregnant woman with COVID-19.* Ultrasound in Obstetrics & Gynecology, 2020.
55. Lütje, S., et al., *Nuclear medicine in SARS-CoV-2 pandemia: 18F-FDG-PET/CT to visualize COVID-19.* Nuklearmedizin, 2020.
56. Deng, Y., et al., *The potential added value of FDG PET/CT for COVID-19 pneumonia.* European Journal of Nuclear Medicine and Molecular Imaging, 2020: p. 1-2.
57. Qin, C., et al., *18 F-FDG PET/CT findings of COVID-19: a series of four highly suspected cases.* European Journal of Nuclear Medicine and Molecular Imaging, 2020: p. 1-6.
58. Zou, S. and X. Zhu, *FDG PET/CT of COVID-19.* Radiology, 2020: p. 200770.
59. Polverari, G., et al., *18F-FDG uptake in Asymptomatic SARS-CoV-2 (COVID-19) patient, referred to PET/CT for Non-Small Cells Lung Cancer restaging.* Journal of Thoracic Oncology, 2020.
60. Joob, B. and V. Wiwanitkit, *18F-FDG PET/CT and COVID-19.* European Journal of Nuclear Medicine and Molecular Imaging, 2020: p. 1-1.
61. Guedj, E., A. Verger, and S. Cammilleri, *PET imaging of COVID-19: the target and the number.* European Journal of Nuclear Medicine and Molecular Imaging, 2020: p. 1-2.
62. Tulchinsky, M., J.S. Fotos, and E. Slonimsky, *Incidental CT Findings Suspicious for Covid-19 Associated Pneumonia on Nuclear Medicine Exams: Recognition and Management Plan.* Clinical Nuclear Medicine, 2020.
63. Poyiadji, N., et al., *COVID-19–associated acute hemorrhagic necrotizing encephalopathy: CT and MRI features.* Radiology, 2020: p. 201187.
64. Rotzinger, D., et al., *Pulmonary embolism in patients with COVID-19: Time to change the paradigm of computed tomography.* Thrombosis research, 2020.
65. Li, D., et al., *False-negative results of real-time reverse-transcriptase polymerase chain reaction for severe acute respiratory syndrome coronavirus 2: role of deep-learning-based CT diagnosis and insights from two cases.* Korean journal of radiology, 2020. **21**(4): p. 505-508.
66. Alimadadi, A., et al., *Artificial intelligence and machine learning to fight COVID-19*. 2020, American Physiological Society Bethesda, MD.
67. Ilyas, M., H. Rehman, and A. Nait-ali, *Detection of Covid-19 From Chest X-ray Images Using Artificial Intelligence: An Early Review.* arXiv preprint arXiv:2004.05436, 2020.
68. Rahmatizadeh, S., S. Valizadeh-Haghi, and A. Dabbagh, *The role of Artificial Intelligence in Management of Critical COVID-19 patients.* Journal of Cellular & Molecular Anesthesia, 2020. **5**(1): p. 16-22.

69. Shi, F., et al., *Review of artificial intelligence techniques in imaging data acquisition, segmentation and diagnosis for covid-19.* IEEE Reviews in Biomedical Engineering, 2020.
70. Ulhaq, A., et al., *Computer Vision for COVID-19 Control: A Survey.* arXiv preprint arXiv:2004.09420, 2020.
71. Nanni, L., S. Ghidoni, and S. Brahnam, *Handcrafted vs. non-handcrafted features for computer vision classification.* Pattern Recognition, 2017. **71**: p. 158-172.
72. LeCun, Y. and Y. Bengio, *Convolutional networks for images, speech, and time series.* The handbook of brain theory and neural networks, 1995. **3361**(10): p. 1995.
73. Gu, J., et al., *Recent advances in convolutional neural networks.* Pattern Recognition, 2018. **77**: p. 354-377.
74. Cohen, J.P., P. Morrison, and L. Dao, *COVID-19 image data collection.* arXiv preprint arXiv:2003.11597, 2020.
75. Zhao, J., et al., *COVID-CT-Dataset: a CT scan dataset about COVID-19.* arXiv preprint arXiv:2003.13865, 2020.
76. Shorten, C. and T.M. Khoshgoftaar, *A survey on image data augmentation for deep learning.* Journal of Big Data, 2019. **6**(1): p. 60.
77. Weiss, K., T.M. Khoshgoftaar, and D. Wang, *A survey of transfer learning.* Journal of Big data, 2016. **3**(1): p. 9.
78. Litjens, G., et al., *A survey on deep learning in medical image analysis.* Medical image analysis, 2017. **42**: p. 60-88.
79. Varshni, D., et al. *Pneumonia Detection Using CNN based Feature Extraction*. in *2019 IEEE International Conference on Electrical, Computer and Communication Technologies (ICECCT)*. 2019. IEEE.
80. Chollet, F., *Xception: deep learning with separable convolutions. 1–14.* arXiv preprint arXiv:1610.02357, 2016.
81. Simonyan, K. and A. Zisserman, *Very deep convolutional networks for large-scale image recognition.* arXiv preprint arXiv:1409.1556, 2014.
82. He, K., et al. *Deep residual learning for image recognition*. in *Proceedings of the IEEE conference on computer vision and pattern recognition*. 2016.
83. Huang, G., et al. *Densely connected convolutional networks*. in *Proceedings of the IEEE conference on computer vision and pattern recognition*. 2017.
84. Narin, A., C. Kaya, and Z. Pamuk, *Automatic detection of coronavirus disease (covid-19) using x-ray images and deep convolutional neural networks.* arXiv preprint arXiv:2003.10849, 2020.
85. Hemdan, E.E.-D., M.A. Shouman, and M.E. Karar, *Covidx-net: A framework of deep learning classifiers to diagnose covid-19 in x-ray images.* arXiv preprint arXiv:2003.11055, 2020.
86. Ghoshal, B. and A. Tucker, *Estimating uncertainty and interpretability in deep learning for coronavirus (COVID-19) detection.* arXiv preprint arXiv:2003.10769, 2020.
87. Khalifa, N.E.M., et al., *Detection of Coronavirus (COVID-19) Associated Pneumonia based on Generative Adversarial Networks and a Fine-Tuned Deep Transfer Learning Model using Chest X-ray Dataset.* arXiv preprint arXiv:2004.01184, 2020.
88. Loey, M., F. Smarandache, and N.E.M. Khalifa, *A Deep Transfer Learning Model with Classical Data Augmentation and CGAN to Detect COVID-19 from Chest CT Radiography Digital Images.* 2020.
89. Farooq, M. and A. Hafeez, *Covid-resnet: A deep learning framework for screening of covid19 from radiographs.* arXiv preprint arXiv:2003.14395, 2020.
90. Apostolopoulos, I.D., S.I. Aznaouridis, and M.A. Tzani, *Extracting possibly representative COVID-19 Biomarkers from X-Ray images with Deep Learning approach and image data related to Pulmonary Diseases.* Journal of Medical and Biological Engineering, 2020: p. 1.
91. Zheng, C., et al., *Deep learning-based detection for COVID-19 from chest CT using weak label.* medRxiv, 2020.

92. Hu, S., et al., *Weakly Supervised Deep Learning for COVID-19 Infection Detection and Classification from CT Images.* arXiv preprint arXiv:2004.06689, 2020.
93. Wang, L. and A. Wong, *COVID-Net: A tailored deep convolutional neural network design for detection of COVID-19 cases from chest radiography images.* arXiv preprint arXiv:2003.09871, 2020.
94. Oh, Y., S. Park, and J.C. Ye, *Deep Learning COVID-19 Features on CXR using Limited Training Data Sets.* IEEE Transactions on Medical Imaging, 2020.
95. Barstugan, M., U. Ozkaya, and S. Ozturk, *Coronavirus (covid-19) classification using ct images by machine learning methods.* arXiv preprint arXiv:2003.09424, 2020.
96. Tang, Z., et al., *Severity assessment of coronavirus disease 2019 (COVID-19) using quantitative features from chest CT images.* arXiv preprint arXiv:2003.11988, 2020.
97. Al-Karawi, D., et al., *Machine Learning Analysis of Chest CT Scan Images as a Complementary Digital Test of Coronavirus (COVID-19) Patients.* medRxiv, 2020.
98. Ai, T., et al., *Correlation of chest CT and RT-PCR testing in coronavirus disease 2019 (COVID-19) in China: a report of 1014 cases.* Radiology, 2020: p. 200642.
99. Yang, Z., et al., *Predictors for imaging progression on chest CT from coronavirus disease 2019 (COVID-19) patients.* Aging (Albany NY), 2020. **12**(7): p. 6037.
100. Colombi, D., et al., *Well-aerated Lung on Admitting Chest CT to Predict Adverse Outcome in COVID-19 Pneumonia.* Radiology, 2020: p. 201433.
101. Bai, H.X., et al., *Performance of radiologists in differentiating COVID-19 from viral pneumonia on chest CT.* Radiology, 2020: p. 200823.
102. Caruso, D., et al., *Chest CT features of COVID-19 in Rome, Italy.* Radiology, 2020: p. 201237.
103. Fan, N., et al., *Imaging characteristics of initial chest computed tomography and clinical manifestations of patients with COVID-19 pneumonia.* Japanese Journal of Radiology, 2020: p. 1-6.
104. Li, X., et al., *CT imaging changes of corona virus disease 2019 (COVID-19): a multi-center study in Southwest China.* Journal of Translational Medicine, 2020. **18**: p. 1-8.
105. Wu, J., et al., *Novel coronavirus pneumonia (COVID-19) CT distribution and sign features.* Zhonghua jie he he hu xi za zhi= Zhonghua Jiehe he Huxi Zazhi= Chinese Journal of Tuberculosis and Respiratory Diseases, 2020. **43**: p. E030-E030.
106. Bernheim, A., et al., *Chest CT findings in coronavirus disease-19 (COVID-19): relationship to duration of infection.* Radiology, 2020: p. 200463.
107. Zhang, R., et al., *CT features of SARS-CoV-2 pneumonia according to clinical presentation: a retrospective analysis of 120 consecutive patients from Wuhan city.* European radiology, 2020: p. 1-10.
108. Zhao, W., et al., *CT scans of patients with 2019 novel coronavirus (COVID-19) pneumonia.* Theranostics, 2020. **10**(10): p. 4606.
109. Han, R., et al., *Early clinical and CT manifestations of coronavirus disease 2019 (COVID-19) pneumonia.* American Journal of Roentgenology, 2020: p. 1-6.
110. Zhao, W., et al., *Relation between chest CT findings and clinical conditions of coronavirus disease (COVID-19) pneumonia: a multicenter study.* American Journal of Roentgenology, 2020. **214**(5): p. 1072-1077.
111. Huang, Z., et al., *Imaging features and mechanisms of novel coronavirus pneumonia (COVID-19): Study Protocol Clinical Trial (SPIRIT Compliant).* Medicine, 2020. **99**(16): p. e19900.
112. Wang, Y., et al., *Temporal changes of CT findings in 90 patients with COVID-19 pneumonia: a longitudinal study.* Radiology, 2020: p. 200843.
113. Xu, X., et al., *Imaging and clinical features of patients with 2019 novel coronavirus SARS-CoV-2.* European journal of nuclear medicine and molecular imaging, 2020: p. 1-6.
114. Liang, T., et al., *Evolution of CT findings in patients with mild COVID-19 pneumonia.* European Radiology, 2020: p. 1-9.

115. Li, K., et al., *The clinical and chest CT features associated with severe and critical COVID-19 pneumonia.* Investigative radiology, 2020. **55**(6): p. 327-331.
116. Wu, J., et al., *Chest CT findings in patients with coronavirus disease 2019 and its relationship with clinical features.* Investigative radiology, 2020. **55**(5): p. 257-261.
117. Liu, K.-C., et al., *CT manifestations of coronavirus disease-2019: a retrospective analysis of 73 cases by disease severity.* European journal of radiology, 2020: p. 108941.
118. Zhong, Q., et al., *CT imaging features of patients with different clinical types of coronavirus disease 2019 (COVID-19).* Journal of Zhejiang University (Medical Science), 2020. **49**(1): p. 0-0.
119. Zhou, Z., et al., *Coronavirus disease 2019: initial chest CT findings.* European Radiology, 2020: p. 1-9.
120. Zhou, S., et al., *CT features of coronavirus disease 2019 (COVID-19) pneumonia in 62 patients in Wuhan, China.* American Journal of Roentgenology, 2020: p. 1-8.
121. Meng, H., et al., *CT imaging and clinical course of asymptomatic cases with COVID-19 pneumonia at admission in Wuhan, China.* Journal of Infection, 2020.
122. Li, Y. and L. Xia, *Coronavirus disease 2019 (COVID-19): role of chest CT in diagnosis and management.* American Journal of Roentgenology, 2020: p. 1-7.
123. Wang, J., et al., *Dynamic changes of chest CT imaging in patients with corona virus disease-19 (COVID-19).* Journal of Zhejiang University (Medical Science), 2020. **49**(1): p. 0-0.
124. Lyu, P., et al., *The performance of chest CT in evaluating the clinical severity of COVID-19 pneumonia: identifying critical cases based on CT characteristics.* Investigative Radiology, 2020.
125. Fang, Y., et al., *Sensitivity of chest CT for COVID-19: comparison to RT-PCR.* Radiology, 2020: p. 200432.
126. Xu, Y.-H., et al., *Clinical and computed tomographic imaging features of novel coronavirus pneumonia caused by SARS-CoV-2.* Journal of Infection, 2020.
127. Lei, P., et al., *The progression of computed tomographic (CT) images in patients with coronavirus disease (COVID-19) pneumonia: Running title: The CT progression of COVID-19 pneumonia.* Journal of Infection, 2020.
128. Yang, S., et al., *Clinical and CT features of early stage patients with COVID-19: a retrospective analysis of imported cases in Shanghai, China.* European Respiratory Journal, 2020. **55**(4).
129. Long, C., et al., *Diagnosis of the Coronavirus disease (COVID-19): rRT-PCR or CT?* European journal of radiology, 2020: p. 108961.
130. Liu, R., et al., *CT imaging analysis of 33 cases with the 2019 novel coronavirus infection.* Zhonghua yi xue za zhi, 2020. **100**(13): p. 1007-1011.
131. Cheng, Z., et al., *Clinical features and chest CT manifestations of coronavirus disease 2019 (COVID-19) in a single-center study in Shanghai, China.* American Journal of Roentgenology, 2020: p. 1-6.
132. Yuan, M., et al., *Association of radiologic findings with mortality of patients infected with 2019 novel coronavirus in Wuhan, China.* PLoS One, 2020. **15**(3): p. e0230548.
133. Dane, B., et al., *Unexpected Findings of Coronavirus Disease (COVID-19) at the Lung Bases on Abdominopelvic CT.* American Journal of Roentgenology, 2020: p. 1-4.
134. Himoto, Y., et al., *Diagnostic performance of chest CT to differentiate COVID-19 pneumonia in non-high-epidemic area in Japan.* Japanese Journal of Radiology, 2020: p. 1.
135. Chung, M., et al., *CT imaging features of 2019 novel coronavirus (2019-nCoV).* Radiology, 2020. **295**(1): p. 202-207.
136. Pan, F., et al., *Time course of lung changes on chest CT during recovery from 2019 novel coronavirus (COVID-19) pneumonia.* Radiology, 2020: p. 200370.
137. Zhu, Z., et al., *Comparison of heart failure and 2019 novel coronavirus pneumonia in chest CT features and clinical characteristics.* Zhonghua xin xue Guan Bing za zhi, 2020. **48**: p. E007-E007.

138. Lei, P., et al., *Clinical and computed tomographic (CT) images characteristics in the patients with COVID-19 infection: What should radiologists need to know?* Journal of X-Ray Science and Technology, 2020(Preprint): p. 1-13.
139. Chate, R.C., et al., *Presentation of pulmonary infection on CT in COVID-19: initial experience in Brazil.* Jornal Brasileiro de Pneumologia, 2020. **46**(2).
140. Zhu, Y., et al., *Clinical and CT imaging features of 2019 novel coronavirus disease (COVID-19).* The Journal of infection, 2020.
141. Lu, T. and H. Pu, *Computed Tomography Manifestations of 5 Cases of the Novel Coronavirus Disease 2019 (COVID-19) Pneumonia From Patients Outside Wuhan.* Journal of thoracic imaging, 2020. **35**(3): p. W90-W93.
142. Xie, X., et al., *Chest CT for typical 2019-nCoV pneumonia: relationship to negative RT-PCR testing.* Radiology, 2020: p. 200343.
143. McGinnis, G.J., et al., *Rapid detection of asymptomatic COVID-19 by CT image-guidance for stereotactic ablative radiotherapy.* Journal of Thoracic Oncology, 2020.
144. Yan, K., et al., *CT challenges the result of SARS-CoV-2 nucleic acid test in a suspected COVID-19 case.* Infection Control & Hospital Epidemiology, 2020: p. 1-5.
145. Qi, X., et al., *CT imaging of coronavirus disease 2019 (COVID-19): from the qualitative to quantitative.* Annals of Translational Medicine, 2020. **8**(5).
146. Zhang, F.-Y., Y. Qiao, and H. Zhang, *CT imaging of the COVID-19.* Journal of the Formosan Medical Association, 2020.
147. Tenda, E.D., et al., *The Importance of Chest CT Scan in COVID-19.* Acta Medica Indonesiana, 2020. **52**(1): p. 68-73.
148. Erturk, S.M., *CT Is Not a Screening Tool for Coronavirus Disease (COVID-19) Pneumonia.* AJR. American Journal of Roentgenology, 2020: p. W1-W1.
149. Xu, R., et al., *CT imaging of one extended family cluster of corona virus disease 2019 (COVID-19) including adolescent patients and "silent infection".* Quantitative Imaging in Medicine and Surgery, 2020. **10**(3): p. 800.
150. Liu, J., H. Yu, and S. Zhang, *The indispensable role of chest CT in the detection of coronavirus disease 2019 (COVID-19).* European journal of nuclear medicine and molecular imaging, 2020.
151. Li, M., et al., *Necessitating repeated chest CT in COVID-19 pneumonia.* Journal of the Formosan Medical Association, 2020.
152. Ufuk, F., *3D CT of Novel Coronavirus (COVID-19) Pneumonia.* Radiology, 2020: p. 201183.
153. Hamer, O.W., et al. *CT morphology of COVID-19: Case report and review of literature*. in *RöFo-Fortschritte auf dem Gebiet der Röntgenstrahlen und der bildgebenden Verfahren*. 2020. © Georg Thieme Verlag KG.
154. Kang, Z., X. Li, and S. Zhou, *Recommendation of low-dose CT in the detection and management of COVID-2019*. 2020, Springer.
155. Dai, W.-c., et al., *CT imaging and differential diagnosis of COVID-19.* Canadian Association of Radiologists Journal, 2020. **71**(2): p. 195-200.
156. Lin, C., et al., *Asymptomatic novel coronavirus pneumonia patient outside Wuhan: The value of CT images in the course of the disease.* Clinical imaging, 2020. **63**: p. 7-9.
157. Lee, E.Y., M.-Y. Ng, and P.-L. Khong, *COVID-19 pneumonia: what has CT taught us?* The Lancet Infectious Diseases, 2020. **20**(4): p. 384-385.
158. Kim, H., *Outbreak of novel coronavirus (COVID-19): What is the role of radiologists?* 2020, Springer.
159. Zhang, X., et al., *CT image of novel coronavirus pneumonia: a case report.* Japanese Journal of Radiology, 2020: p. 1-2.
160. Singh, N. and J. Fratesi, *Chest CT imaging of an early Canadian case of COVID-19 in a 28-year-old man.* CMAJ, 2020. **192**(17): p. E455-E455.
161. Tsou, I., et al., *Planning and coordination of the radiological response to the coronavirus disease 2019 (COVID-19) pandemic: the Singapore experience.* Clinical Radiology, 2020.

162. Vu, D., et al., *Three unsuspected CT diagnoses of COVID-19.* Emergency Radiology, 2020: p. 1-4.
163. Çinkooğlu, A., S. Bayraktaroğlu, and R. Savaş, *Lung changes on chest CT during 2019 novel coronavirus (COVID-19) pneumonia.* European Journal of Breast Health, 2020. **16**(2): p. 89.
164. Asadollahi-Amin, A., et al., *Lung Involvement Found on Chest CT Scan in a Pre-Symptomatic Person with SARS-CoV-2 Infection: A Case Report.* Tropical medicine and infectious disease, 2020. **5**(2): p. 56.
165. Chen, X., et al., *Dynamic Chest CT Evaluation in Three Cases of 2019 Novel Coronavirus Pneumonia.* Archives of Iranian Medicine, 2020. **23**(4): p. 277-280.
166. Hu, X., et al., *CT imaging of two cases of one family cluster 2019 novel coronavirus (2019-nCoV) pneumonia: inconsistency between clinical symptoms amelioration and imaging sign progression.* Quantitative Imaging in Medicine and Surgery, 2020. **10**(2): p. 508.
167. Lim, Z.Y., et al., *Variable computed tomography appearances of COVID-19.* Singapore medical journal, 2020.
168. Ostad, S., S. Haseli, and P. Iranpour, *CT Manifestation of COVID-19 Pneumonia; Role of Multiplanar Imaging.* Academic Radiology, 2020.
169. Qanadli, S.D., C. Beigelman-Aubry, and D.C. Rotzinger, *Vascular Changes Detected With Thoracic CT in Coronavirus Disease (COVID-19) Might Be Significant Determinants for Accurate Diagnosis and Optimal Patient Management.* AJR. American journal of roentgenology, 2020: p. W1.
170. Mungmungpuntipantip, R. and V. Wiwanitkit, *Clinical Features and Chest CT Manifestations of Coronavirus Disease (COVID-19).* American Journal of Roentgenology, 2020. **215**(7): p. W1-W1.
171. Danrad, R., D.L. Smith, and E.K. Kerut, *Radiologic Chest CT Findings From COVID-19 in Orleans Parish, Lousiana.* Echocardiography, 2020.
172. Joob, B. and V. Wiwanitkit, *Chest CT Findings and Clinical Conditions of Coronavirus Disease (COVID-19).* AJR. American journal of roentgenology, 2020: p. W1-W1.
173. Li, C.-X., et al., *Clinical Study and CT Findings of a Familial Cluster of Pneumonia with Coronavirus Disease 2019 (COVID-19).* Sichuan da xue xue bao. Yi xue ban= Journal of Sichuan University. Medical Science Edition, 2020. **51**(2): p. 155-158.
174. Guan, W., J. Liu, and C. Yu, *CT Findings of Coronavirus Disease (COVID-19) Severe Pneumonia.* American Journal of Roentgenology, 2020. **214**(5): p. W85-W86.
175. Li, M., et al., *Coronavirus disease (COVID-19): spectrum of CT findings and temporal progression of the disease.* Academic radiology, 2020.
176. Lei, P., B. Fan, and Y. Yuan, *The evolution of CT characteristics in the patients with COVID-19 pneumonia.* Journal of Infection, 2020.
177. Gross, A., et al. *CT appearance of severe, laboratory-proven coronavirus disease 2019 (COVID-19) in a Caucasian patient in Berlin, Germany*. in *RöFo-Fortschritte auf dem Gebiet der Röntgenstrahlen und der bildgebenden Verfahren*. 2020. © Georg Thieme Verlag KG.
178. Shi, F., et al., *2019 Novel Coronavirus (COVID-19) Pneumonia with Hemoptysis as the Initial Symptom: CT and Clinical Features.* Korean Journal of Radiology, 2020. **21**(5): p. 537.
179. Qu, J., et al., *Atypical lung feature on chest CT in a lung adenocarcinoma cancer patient infected with COVID-19.* Annals of Oncology, 2020.
180. An, P., et al., *CT manifestations of novel coronavirus pneumonia: a case report.* Balkan Medical Journal, 2020. **37**(3): p. 163.
181. Wei, J., et al., *2019 novel coronavirus (COVID-19) pneumonia: serial computed tomography findings.* Korean Journal of Radiology, 2020. **21**(4): p. 501-504.
182. Fang, X., et al., *Changes of CT Findings in a 2019 Novel Coronavirus (2019-nCoV) pneumonia patient.* QJM: An International Journal of Medicine, 2020. **113**(4): p. 271-272.
183. Duan, Y.-n. and J. Qin, *Pre-and posttreatment chest CT findings: 2019 novel coronavirus (2019-nCoV) pneumonia.* Radiology, 2020. **295**(1): p. 21-21.

184. Shi, H., X. Han, and C. Zheng, *Evolution of CT manifestations in a patient recovered from 2019 novel coronavirus (2019-nCoV) pneumonia in Wuhan, China.* Radiology, 2020. **295**(1): p. 20-20.
185. Fang, Y., et al., *CT manifestations of two cases of 2019 novel coronavirus (2019-nCoV) pneumonia.* Radiology, 2020. **295**(1): p. 208-209.
186. Kanne, J.P., *Chest CT findings in 2019 novel coronavirus (2019-nCoV) infections from Wuhan, China: key points for the radiologist*. 2020, Radiological Society of North America.
187. Adair II, L.B. and E.J. Ledermann, *Chest CT findings of early and progressive phase COVID-19 infection from a US patient.* Radiology Case Reports, 2020.
188. Burhan, E., et al., *Clinical Progression of COVID-19 Patient with Extended Incubation Period, Delayed RT-PCR Time-to-positivity, and Potential Role of Chest CT-scan.* Acta Medica Indonesiana, 2020. **52**(1): p. 80.
189. Feng, H., et al., *A case report of COVID-19 with false negative RT-PCR test: necessity of chest CT.* Japanese Journal of Radiology, 2020: p. 1-2.
190. Hao, W. and M. Li, *Clinical diagnostic value of CT imaging in COVID-19 with multiple negative RT-PCR testing.* Travel medicine and infectious disease, 2020.
191. Yang, W. and F. Yan, *Patients with RT-PCR-confirmed COVID-19 and Normal Chest CT.* Radiology, 2020. **295**(2): p. E3-E3.
192. Lei, J., et al., *CT imaging of the 2019 novel coronavirus (2019-nCoV) pneumonia.* Radiology, 2020. **295**(1): p. 18-18.
193. Apostolopoulos, I.D. and T.A. Mpesiana, *Covid-19: automatic detection from x-ray images utilizing transfer learning with convolutional neural networks.* Physical and Engineering Sciences in Medicine, 2020: p. 1.
194. Zhang, J., et al., *Covid-19 screening on chest x-ray images using deep learning based anomaly detection.* arXiv preprint arXiv:2003.12338, 2020.
195. Sethy, P.K. and S.K. Behera, *Detection of coronavirus disease (covid-19) based on deep features.* Preprints, 2020. **2020030300**: p. 2020.
196. Abbas, A., M.M. Abdelsamea, and M.M. Gaber, *Classification of COVID-19 in chest X-ray images using DeTraC deep convolutional neural network.* arXiv preprint arXiv:2003.13815, 2020.
197. Afshar, P., et al., *Covid-caps: A capsule network-based framework for identification of covid-19 cases from x-ray images.* arXiv preprint arXiv:2004.02696, 2020.
198. Chowdhury, M.E., et al., *Can AI help in screening viral and COVID-19 pneumonia?* arXiv preprint arXiv:2003.13145, 2020.
199. Li, X. and D. Zhu, *Covid-xpert: An ai powered population screening of covid-19 cases using chest radiography images.* arXiv preprint arXiv:2004.03042, 2020.
200. Karim, M., et al., *Deepcovidexplainer: Explainable covid-19 predictions based on chest x-ray images.* arXiv preprint arXiv:2004.04582, 2020.
201. Hall, L.O., et al., *Finding covid-19 from chest x-rays using deep learning on a small dataset.* arXiv preprint arXiv:2004.02060, 2020.
202. Rajaraman, S., et al., *Iteratively Pruned Deep Learning Ensembles for COVID-19 Detection in Chest X-rays.* arXiv preprint arXiv:2004.08379, 2020.
203. Luz, E., et al., *Towards an Efficient Deep Learning Model for COVID-19 Patterns Detection in X-ray Images.* arXiv preprint arXiv:2004.05717, 2020.
204. Tartaglione, E., et al., *Unveiling COVID-19 from Chest X-ray with deep learning: a hurdles race with small data.* arXiv preprint arXiv:2004.05405, 2020.
205. Ezzat, D. and H.A. Ella, *GSA-DenseNet121-COVID-19: a Hybrid Deep Learning Architecture for the Diagnosis of COVID-19 Disease based on Gravitational Search Optimization Algorithm.* arXiv preprint arXiv:2004.05084, 2020.
206. Hammoudi, K., et al., *Deep Learning on Chest X-ray Images to Detect and Evaluate Pneumonia Cases at the Era of COVID-19.* arXiv preprint arXiv:2004.03399, 2020.

207. Kumar, R., et al., *Accurate Prediction of COVID-19 using Chest X-Ray Images through Deep Feature Learning model with SMOTE and Machine Learning Classifiers.* medRxiv, 2020.
208. Khan, A.I., J.L. Shah, and M. Bhat, *CoroNet: A Deep Neural Network for Detection and Diagnosis of Covid-19 from Chest X-ray Images.* arXiv preprint arXiv:2004.04931, 2020.
209. Santosh, K., D. Das, and U. Pal, *Truncated Inception Net: COVID-19 Outbreak Screening using Chest X-rays.* 2020.
210. Khobahi, S., C. Agarwal, and M. Soltanalian, *CoroNet: A Deep Network Architecture for Semi-Supervised Task-Based Identification of COVID-19 from Chest X-ray Images.* medRxiv, 2020.
211. Rahimzadeh, M. and A. Attar, *A New Modified Deep Convolutional Neural Network for Detecting COVID-19 from X-ray Images.* arXiv preprint arXiv:2004.08052, 2020.
212. Loey, M., F. Smarandache, and N.E.M. Khalifa, *Within the Lack of COVID-19 Benchmark Dataset: A Novel GAN with Deep Transfer Learning for Corona-virus Detection in Chest X-ray Images.* 2020.
213. Pereira, R.M., et al., *COVID-19 identification in chest X-ray images on flat and hierarchical classification scenarios.* Computer Methods and Programs in Biomedicine, 2020: p. 105532.
214. Gozes, O., et al., *Rapid ai development cycle for the coronavirus (covid-19) pandemic: Initial results for automated detection & patient monitoring using deep learning ct image analysis.* arXiv preprint arXiv:2003.05037, 2020.
215. Shan, F., et al., *Lung infection quantification of covid-19 in ct images with deep learning.* arXiv preprint arXiv:2003.04655, 2020.
216. Butt, C., et al., *Deep learning system to screen coronavirus disease 2019 pneumonia.* Applied Intelligence, 2020: p. 1.
217. Wang, S., et al., *A deep learning algorithm using CT images to screen for Corona Virus Disease (COVID-19).* MedRxiv, 2020.
218. Li, L., et al., *Artificial intelligence distinguishes COVID-19 from community acquired pneumonia on chest CT.* Radiology, 2020: p. 200905.
219. Huang, L., et al., *Serial quantitative chest CT assessment of COVID-19: deep-learning approach.* Radiology: Cardiothoracic Imaging, 2020. **2**(2): p. e200075.
220. Song, Y., et al., *Deep learning enables accurate diagnosis of novel coronavirus (COVID-19) with CT images.* medRxiv, 2020.
221. Gozes, O., et al., *Coronavirus detection and analysis on chest ct with deep learning.* arXiv preprint arXiv:2004.02640, 2020.
222. Chen, J., et al., *Deep learning-based model for detecting 2019 novel coronavirus pneumonia on high-resolution computed tomography: a prospective study.* medRxiv, 2020.
223. Chaganti, S., et al., *Quantification of tomographic patterns associated with COVID-19 from chest CT.* arXiv preprint arXiv:2004.01279, 2020.
224. He, X., et al., *Sample-Efficient Deep Learning for COVID-19 Diagnosis Based on CT Scans.* medRxiv, 2020.
225. Fu, M., et al., *Deep Learning-Based Recognizing COVID-19 and other Common Infectious Diseases of the Lung by Chest CT Scan Images.* medRxiv, 2020.
226. Chen, X., L. Yao, and Y. Zhang, *Residual Attention U-Net for Automated Multi-Class Segmentation of COVID-19 Chest CT Images.* arXiv preprint arXiv:2004.05645, 2020.
227. Zhou, T., S. Canu, and S. Ruan, *An automatic COVID-19 CT segmentation based on U-Net with attention mechanism.* arXiv preprint arXiv:2004.06673, 2020.
228. Mobiny, A., et al., *Radiologist-Level COVID-19 Detection Using CT Scans with Detail-Oriented Capsule Networks.* arXiv preprint arXiv:2004.07407, 2020.
229. Wu, Y.-H., et al., *JCS: An Explainable COVID-19 Diagnosis System by Joint Classification and Segmentation.* arXiv preprint arXiv:2004.07054, 2020.
230. Maghdid, H.S., et al., *Diagnosing COVID-19 pneumonia from X-ray and CT images using deep learning and transfer learning algorithms.* arXiv preprint arXiv:2004.00038, 2020.

231. Alom, M.Z., et al., *COVID_MTNet: COVID-19 Detection with Multi-Task Deep Learning Approaches.* arXiv preprint arXiv:2004.03747, 2020.
232. Razzak, I., et al., *Improving Coronavirus (COVID-19) Diagnosis using Deep Transfer Learning.* medRxiv, 2020.